\documentclass[conference]{IEEEtran}

\usepackage{amsmath}
\usepackage{amsthm}

\usepackage{algorithm}
\usepackage{algorithmicx}
\usepackage{algpseudocode}

\usepackage[tight,footnotesize]{subfigure}
\usepackage{url}
\usepackage{booktabs}
\usepackage[utf8x]{inputenc}
\usepackage{graphicx}
\usepackage{multirow}
\usepackage{rotating}
\usepackage{todonotes}
\usepackage{amssymb}
\usepackage{listings}

\newtheorem{definition}{Definition}
\newtheorem{lemma}{Lemma}[section]

\newtheorem{theorem}{Theorem}[section]

\algrenewcommand{\alglinenumber}[1]{\algprintlinenumber{#1}:}
\algnewcommand{\algprintlinenumber}[1]{\footnotesize#1}
\newcommand{\myref}[1]{\hyperref[#1]{\algprintlinenumber{\ref*{#1}}}}

\newcommand{\remove}[1]{}

\algdef{SE}[DOWHILE]{Do}{doWhile}{\algorithmicdo}[1]{\algorithmicwhile\ #1}%

\title{A Non-blocking Buddy System for Scalable Memory Allocation on Multi-core Machines}

\author{
\IEEEauthorblockN{Romolo Marotta, Mauro Ianni, Andrea Scarselli, Alessandro Pellegrini}
\IEEEauthorblockA{Sapienza, University of Rome}
\and
\IEEEauthorblockN{Francesco Quaglia}
\IEEEauthorblockA{University of Rome ``Tor Vergata''}
}

 \definecolor{nc}{rgb}{0.9,0.8,0.5}

\usepackage{tcolorbox}
\tcbset{width=\linewidth,boxrule=0pt,colback=nc,arc=0pt,auto outer arc,left=0pt,right=0pt,boxsep=0pt}
\usepackage{xcolor}
\usepackage[normalem]{ulem}

\newcommand{\algosizefinal}{\footnotesize}

\begin{document}

\maketitle

\begin{abstract}
Common implementations of core memory allocation components, like the Linux buddy system, handle concurrent allocation/release requests by synchronizing threads via spin-locks.
This approach is not prone to scale with large thread counts, a problem that has been addressed in the literature by introducing layered allocation services or replicating the core allocators---the bottom most ones within the layered architecture.
Both these solutions tend to reduce the pressure of actual concurrent accesses to each individual core allocator.
In this article we explore an alternative approach to scalability of memory allocation/release, which can be still combined with those literature proposals.
We present a fully non-blocking buddy-system, where threads performing concurrent allocations/releases do not undergo any spin-lock based synchronization.
Our solution 
allows threads to proceed in parallel, and commit their allocations/releases unless a conflict is materialized while handling its metadata.
Conflict detection relies on conventional atomic machine instructions in the Read-Modify-Write (RMW) class.
Beyond improving scalability and performance, our solution can  also avoid wasting clock cycles for spin-lock operations by threads that could in principle carry out their memory allocation/release in full concurrency.
Thus, it is resilient to performance degradation---in face of concurrent accesses---independently of the current level of fragmentation of the handled memory blocks.

\end{abstract}


\section{Introduction}

In standard libraries or in an Operating System (OS), memory allocation  is de-facto a {\em shared-data management} problem. In fact, allocators deal with the issue of managing memory buffers in face of requests that can be concurrently issued by multiple threads.
This requires thread-coordination mechanisms in order to guarantee the coherence of the state of the memory allocator at any time.


A classical coordination approach, which is widely used in allocators, consists in using spin-locks. With this approach, the manipulation of shared data representing the state of the memory allocator is implemented as a critical section, hampering scalability.
This is a relevant issue, since the level of concurrency is increasingly exacerbated because of the modern-hardware trend towards multi/many-core technologies.


The historical approach
for reducing the impact of thread coordination
(and the associated costs) on performance with concurrent memory allocations/releases
is based on either:
 \begin{itemize}
 \item[(a)] pre-reserving {\em arenas}, namely memory segments, for each individual thread---this is what typically happens in user-space allocators \cite{glibc, Pell09}--- or
 \item[(b)] the usage of intermediate allocation services, called \emph{cached allocators}---as for the case of OS-kernel allocation services based on quick-lists \cite{Gorman:2004:ULV:983550}.
 \end{itemize}

Both approaches aim to reduce the likelihood of inducing large volumes of concurrent accesses to the core allocator that is in charge of ultimately delivering memory, either logical or physical.
As for point (a), this is achieved by resorting to the core allocator only when the thread's own pre-reserved arena, which is not accessed by other threads, gets exhausted.
As for point (b), cached allocation diminishes the pressure of concurrent accesses to the core allocator---e.g. a kernel-level buddy system---by having upper-level allocators destined to serve specific memory requests, such as those associated with a given size and/or memory alignment---this is the classical case of kernel-level page-table allocations
or even SLAB allocations.
In such a case, concurrent threads performing memory allocations require coordination only when they need to access the same cached allocator instance
or when this allocator is exhausted and a new
memory segment needs to be taken from the core allocator.


Concurrent allocation/release operations have also been tackled by
creating data separation on the core allocator via multi-instance approaches, and redirecting the requests towards different instances. This further increases the likelihood of saving the requests from actual conflicts that may lead one or more of them to be delayed.
For instance, this approach is used in OS-kernel physical memory management on large scale NUMA (Non-Uniform-Memory-Access) machines, where multiple disjoint instances of the buddy system are included, each one managing physical frames to be allocated from---or released to---different NUMA nodes.

Generally speaking, literature approaches make memory management be extremely layered: the upper the layer, the more it satisfies specific requests.
In this context, we can define \emph{back-end} and \emph{front-end} allocators.
The former, previously denoted as core allocators, handle the lowest level of memory management.
The latter are built on top the back-end allocator with the goal of reducing the pressure of (concurrent) accesses to it and satisfying more specific purposes.

In this article we tackle the issue of scalability of back-end memory allocation, which is an orthogonal approach with respect to reducing the pressure to core allocators by designing front-end allocators, e.g., adopting (a), (b), or multiple-instance approaches.
In particular, our contribution is the design of a non-blocking back-end allocator instance implementing the buddy-system specification, where concurrent allocations/releases are not coordinated via spin-locks.

In our approach, coordination and the guarantee of an always-coherent state of the buddy system are achieved by only relying on individual Read-Modify-Write (RMW) instructions executed along the critical path of allocation/release operations.
These instructions are exploited to detect whether concurrent requests have conflicted on the same portion of the allocator metadata.
This may lead a few of them to be retried as in the classical non-blocking algorithmic paradigm devised in \cite{Her91}.
However, if conflicts do not materialize, then our proposal saves the latency that would be spent by lock-based approaches, which temporarily block concurrent operations a-priori of their execution.


Clearly, our non-blocking solution can be used in combination with any already existing scheme aimed at diminishing the pressure of concurrent accesses to the back-end allocator, e.g. by introducing multiple instances or combining it with front-end allocators.
This is because our unique goal is to provide a memory allocation system that simply optimizes the management of those concurrent accesses with respect to lock-based approaches. On the other hand, having a more efficient back-end allocator can allow to reduce the impact of, e.g., pre-allocation on actual memory unavailability in scenarios where there are skewed memory usages by different threads---so that the pre-reserved memory for a given thread cannot be used for serving a more memory-demanding one---or by different cached allocators.

Our buddy-system implementation has been released as free software\footnote{\url{https://github.com/HPDCS/NBBS}}, and we also provide experimental data demonstrating the actual scalability of our proposal.

The remainder of this article is structured as follows. In Section \ref{related} we discuss related work.
The non-blocking buddy system is presented in Section \ref{buddy}. Experimental data are provided in Section \ref{data}.
For space constraints the proof of safety and progress of our non-blocking buddy system has been removed from this submission. The reader can anyhow refer to the report in \cite{noi-archive}, where we included the proof.

\section{Related Work}
\label{related}




The seminal literature work providing the base-ground for non-blocking coordination in the management of shared-data is \cite{Her91}.
This work has opened the path towards the design and implementation of algorithms that can well fit the scalability requirements imposed by modern (large-scale) multi-core machines.
Also, the avoidance of lock-based coordination in such new algorithmic class has indirectly offered the opportunity to develop coordination algorithms that are more suited for CPU-stealing contexts such as Cloud-based computing.
In these contexts, the de-schedule of a lock-holding thread because of CPU-steals can lead to detrimental performance and waste of energy because of the stretch of the spin-locking phase by other threads attempting to access the same critical section.
Along this direction, a lot of effort has been spent in developing non-blocking versions of classical data structures such as lists or queues \cite{Har01,RamalheteC17a}, hash-tables \cite{PurcellH05}, registers \cite{Ian17,Lar09}
 binary-search trees \cite{Bronson2010,Natarajan2014} and priority queues \cite{Linden2013, MarottaI0Q16}.

In any case, while many solutions have been devised in order to reduce the negative impact of concurrency and synchronization in memory allocation/deallocation by relying on pre-reserving or caching, no one fully faces the problem of concurrent accesses to back-end allocators.
The work in \cite{Dice:2002:MLM:773039.512451} provides non-blocking capabilities of memory allocations with high probability, just depending on the pattern of memory allocations/releases and overall memory usage. This solution is in fact based on pre-reserving memory to be delivered to specific threads (or CPU-cores), and resorts to lock-based coordination across the threads
whenever the pre-reserved memory is fully used by a thread and the global state of the memory allocator needs to be changed in order to provide a new pre-reserved area.
Similar approaches where threads operate on pre-partitioned heaps---hence on different memory-allocator instances---have been presented in \cite{Michael:2004:SLD:996841.996848, LeeKKEJK11, berger}.
Similarly to the previous work, these proposal still does not address the problem of avoiding blocking allocations/releases in scenarios where a same allocator instance can be concurrently accessed by multiple threads. This issue is partially alleviated in \cite{sfmalloc}, where a non-blocking stack data structure is used to post memory across different threads within the pre-reserving scheme.

The solution in \cite{CIT.2010.206} is suited for SIMD systems and exploits worker threads operating in isolation (on different data portions) in order to deliver memory to a pool of requesting threads. The workers do not block each other thanks to pre-partitioned accesses to the allocator data structures---they  actually  operate on different allocators. Hence this approach does not provide a mechanism to perform non-blocking memory allocation/release on a same shared instance of the allocator.
On the other hand, this solution looks to be a reasonably scalable approach for devices such as GPUs where multiple threads typically do the same kind of operations in parallel, such as requesting memory to the worker in charge of managing the allocator---which ultimately delivers a unique memory block, each portion of which is used by a different requesting thread.

We note that our approach is de-facto orthogonal to any approach that tries to improve the scalability of memory allocation via pre-reserving, since we optimize the actual handling of concurrent operations on the core allocator---a buddy system in our case---on top of which pre-reserving can be put in place.
Furthermore, it allows to cope with scenarios where pre-reserving or cached allocation could be not fully adequate.
In particular, cached allocators and multi-instance approaches do not fully cope with 
skews in memory utilization of different caches/instances.
This is the case of OS-kernel physical memory allocation, where requests can be issued by active threads either for the execution of specific system calls, or because of the materialization into physical memory of logical pages upon user-space code accesses to a previously mapped logical memory region---like an {\tt mmap}-ed page on Unix systems.
In these scenarios, the skew of memory requests towards a given instance of allocation service---such as an allocator operating in a given NUMA node selected on the basis of memory-policies associated with the requesting threads---can give rise to a peak of requests
saturating cached allocation and requiring coordinated concurrent accesses to the underlying buddy-system instance.

A solution with goals similar to the ones pointed to by our proposal
can be found in \cite{univis91049556}.
 Here the authors present a concurrent non-blocking memory allocator relying on the so called {\em helping} strategy---where threads help each other trying to avoid blocking scenarios. This solution is based on a kind of {\em conditional atomic dec} instruction which is not currently supported by conventional processors. Conversely, our proposal only relies on machine instructions offered by off-the-shelf CPUs.
 Moreover, as pointed by the same authors, this solution is not able to early detect memory fragmentation, thus possibly leading to a large amount of false positives and unnecessary retries.


%

\section{The Non-blocking Buddy System}
\label{buddy}

Before introducing our solution, a preliminary notion of buddy-system specification is given.
A buddy system is a memory allocator that divides a contiguous memory region into partitions, namely memory blocks,
by splitting recursively it into halves.
These partitions are always contiguous in memory, thus minimizing external fragmentation.
In order to satisfy a memory allocation request, it searches for a memory block large enough to satisfy the request.

Every memory block has an order, which is an integer ranging from 0 to a specified upper limit.
The size of a block of order $n$ is proportional to $2^n$, so that blocks are exactly twice the size of blocks that are at one order lower.

The peculiarity of a buddy system is that memory allocation requests can be satisfied by: i) returning a block if already available; ii) splitting higher-order free blocks into smaller ones; iii) merging lower-order free blocks into a larger one.
The binary arrangement of blocks allows to simplify the split/merge of blocks making them very fast thanks to the exploitation of bit-wise operations.
For example, given the address of a block of a given order, finding the address of the relative buddy is a simple arithmetic operation and does not need de-referencing pointers.

In the following four sections, we will discuss the metadata used in our solution, the non-blocking allocation and release procedures, and finally a relevant optimization of our proposal.

\subsection{Basics}
\label{basics}

Our non-blocking buddy system keeps track of the state of the memory segment used for serving allocations by the means of a static complete binary tree.
The tree has a predefined depth $d$ and its structure is assumed to be already materialized in memory.
The root of the tree corresponds to (and keeps track of the state of) the entire memory segment within which allocations will take place.
Each child of a node represents a portion (a half) of the parent's chunk of memory, while the leaves represent the state of the minimum allocatable memory chunks, called \emph{allocation units}.
In particular, according to the classical buddy-system structure, if a node at level $i$ has size $s$, the children of this node, located at level $i+1$, have size $s/2$, and the union of the blocks of memory associated with the children form a larger block of memory that exactly corresponds to the parent.
This means that, considering an overall size of memory managed by the buddy system equal to ${\tt total\_memory}$, the memory size managed by a node at level $i$
is equal to ${\tt total\_memory}/ 2^i$, and the allocation units (corresponding to the leaves of the tree) have size equal to ${\tt total\_memory} / 2^{\textit{d}}$.

\begin{figure}[t]
\centering
\includegraphics[width=0.7\linewidth]{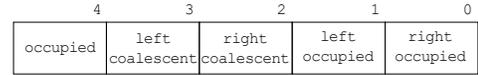}
\caption{Node's status bits.}
\label{fig:stat}
\end{figure}

In our non-blocking buddy system, each node in the tree embeds a bit-mask with 5 relevant bits, organized as in Figure \ref{fig:stat}.
These are used to represent the state of the node itself---thus of the corresponding memory chunk---and of its sub-trees (if any) according to the following semantic: 
\begin{itemize}
\item ${\tt occupied}$, this flag indicates whether an allocation operation has targeted exactly that node, meaning that an allocation request
    has been served by a memory chunk corresponding to that node;
\item 
${\tt left\_occupied}$ and ${\tt right\_occupied}$, these flags indicate if the branches (left and right, respectively) covered by the node
 are totally or partially occupied. This means that
  some allocation request has been served by a node in these sub-trees;
\item ${\tt left\_coalescent}$ and ${\tt right\_coalescent}$, these flags indicate whether a memory release operation is currently in place in any of the two sub-trees. In other words, these two flags indicate whether the node is currently in a transient state because of memory-release modifications running on the relative sub-trees.
\end{itemize}


In order to correctly manipulate the status bits while handling concurrent operations, our solution relies
on \textit{Read-Modify-Write} (RMW) instructions offered by conventional
architectures, like {\sf x86}.
These instructions are able to atomically retrieve a single-word memory value and (conditionally) modify it contextually to the retrieval.
In particular, our operations are based on the \textit{Compare-and-Swap} (CAS) RMW instruction.
This instruction updates a given memory location only if its current value is equal to an input value provided to the instruction; it then returns the success (or the failure) of the operation.

As in common buddy-system organizations, each allocation of a given size $s$ is mapped to its immediately higher size corresponding to ${\tt total\_memory} / 2^i$ with $i \in [0, d]$.
Also, when a node is allocated, its sub-tree is considered to be fully occupied---this is exactly what the {\tt occupied} flag takes care of signaling.
Nevertheless, we do not reflect node occupancy on the lower-level nodes, which helps us saving atomic RMW instructions 
for updating the corresponding status bits.

We represent the tree of nodes keeping track of the buddy-system state with an array of $2^{d+1}-1$ elements, which we refer to as ${\tt tree}$[]. We place the root node at index 1 and exploit the conventional rule that,
given a node with index $n$, the left child of this node is at index $n*2$ and the right child is at index $n*2 + 1$.
This representation fulfills the condition that nodes belonging to the same level of the tree are placed in a contiguous portion of the array, thus simplifying the search for free chunks of a given size while performing allocations.
In fact, starting from the amount of requested memory $s$ it is possible to compute the target level---the one containing nodes useful to serve the request based on its size---as $level = {\lfloor log_2({\tt total\_memory}/{s}) \rfloor}$ upper-bounded by the value $d$.
The nodes belonging to this level are those with index $n \in [2^{level}, 2^{level+1}-1]$.

\begin{figure}[t]
\centering
\includegraphics[width=0.8\linewidth]{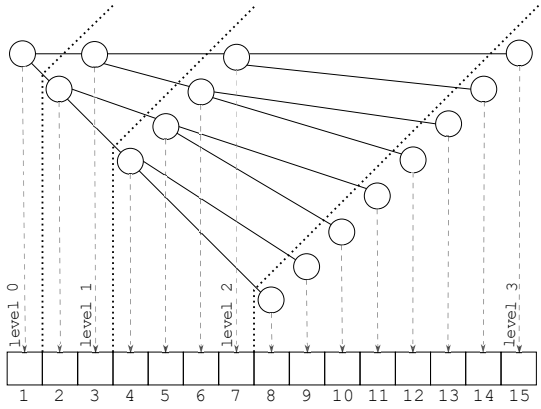}
\caption{Array representation of the tree structure.}
\label{fig:tree}
\end{figure}

A graphical representation of this structure is shown in Figure \ref{fig:tree}.
Denoting with ${\tt base\_address}$ the start of the memory region (either physical or logical) managed by the allocator, for each node with index $n$ it is possible to compute the starting address ${\tt starting_n}$ and the size  ${\tt size_n}$ of the corresponding memory chunk, as well as the level associated with the node, according to the following rules:

\begin{gather}
{\tt level_n} = {\lfloor log_2(n) \rfloor} \label{eq:getlevel}\\
{\tt size_n} = \frac{{\tt total\_memory}}{2^{level_n}}\\
{\tt starting_n} = {\tt base\_address} + (n - 2^{level_n}) * {\tt size_n} \label{eq:getstarting}
\end{gather}

We couple the ${\tt tree}$[] array with another array, called ${\tt index}$[].
This array is used to keep track of the indexes of the nodes that have been used for serving memory requests, and which have not yet been released.
Given the address, we can easily retrieve the index of the node and use this information during a release operation, whose API receives the base address of the to-be-freed node as its unique parameter.

\remove{
\sout{
\sout{Essentially, these are the {\it occupied} nodes.}
Clearly, no more memory requests than the number of leaves can be satisfied by the buddy system at a same time instant.
\sout{In fact, such a number is only reached when standing allocations are all bound to the minimum granularity chunks---say the allocation units---each of which is associated with a different leaf.}
This is the reason why the array ${\tt index}$[] does not require more elements than the number of leaves to keep track of the indexes associated with standing allocations.
At the same time, the index of the entry within the array tells what the memory address is, relative to ${\tt base\_address}$, featuring the memory allocation.
In other words, each entry of this array is implicitly associated---depending on the displacement of the entry---with a memory address of a chunk possibly used to serve a given request.
However, being chunks of different sizes aligned when moving along the left sub-tree of the buddy system, then an entry can keep a different index depending on what is the level of granularity according to which the corresponding chunk (at that memory address) has been used for serving the request.}
}


We present our non-blocking allocation/release algorithms by relying on a few additional notations.
The maximum reachable level in the tree-based organization is stored in a variable denoted as ${\tt depth}$. Moreover, we denote as ${\tt min\_size}$ the variable keeping the size of the allocation units associated with the leaves, and as ${\tt max\_size}$ the variable that keeps track of the maximum amount of
 memory allocatable with a single request (available at level ${\tt max\_level}$)---clearly, ${\tt max\_size} \leq {\tt total\_memory}$.

To extract and manipulate the status bits within the nodes of the tree, the following bit-masks are used:
\begin{flalign}
	{\tt OCC\_RIGHT } & = {\tt 0x1}                                       \nonumber\\
	{\tt OCC\_LEFT  } & = {\tt 0x2}                                       \nonumber\\
	{\tt COAL\_RIGHT} & = {\tt 0x4}                                       \nonumber\\
	{\tt COAL\_LEFT } & = {\tt 0x8}                                       \nonumber\\
	{\tt OCC        } & = {\tt 0x10}                                       \nonumber\\		
	{\tt BUSY       } & = ({\tt OCC}~|~{\tt OCC\_LEFT}~|~{\tt OCC\_RIGHT}) \nonumber
\end{flalign}

Each memory allocation/release operation consists in traversing the tree up to {\tt max\_level} in order to correctly manipulate the status bits of the traversed nodes. In fact, these bits need to be (re)aligned to the new state of the buddy system, depending on the type of operation that is performed. In particular, a memory release operation starts from the node to be released, while a memory allocation operation starts from whichever node at the target level---this depends on the size of the memory allocation request.

Given a value $val$ of the status bits of a given node and the index $child$ of the previous node traversed while moving towards the root, which  is a child of the given node, the following status-bit manipulation functions are offered to simplify the exposition:

\begin{description}
	\item[$clean\_coal(val, child) =$]
	\item[]$val$ $\&$ $\neg({\tt COAL\_LEFT}$ $>>$ $mod_2(child))$
	\item[$mark(val, child) =$]
	\item[]$val$ $\mathrel{|}$ $({\tt OCC\_LEFT}$ $>>$ $mod_2(child))$
	\item[$unmark(val, child) =$]
	\item[]$val$ $\&$ $\neg(({\tt OCC\_LEFT} \mathrel{|} {\tt COAL\_LEFT})$ $>>$ $mod_2(child))$
	\item[$is\_coal(val, child) =$]
	\item[]$val$ $\&$ $({\tt COAL\_LEFT}$ $>>$ $mod_2(child))$
	\item[$is\_occ\_buddy(val, child) =$]
	\item[]$val$ $\&$ $({\tt OCC\_RIGHT}$ $<<$ $mod_2(child))$
	\item[$is\_coal\_buddy(val, child) =$]
	\item[]$val$ $\&$ $({\tt COAL\_RIGHT}$ $<<$ $mod_2(child))$
\end{description}
$clean\_coal(val, child)$ sets to zero the coalescing bit relative to the branch of the child.
 $mark(val, child)$ sets to one the occupancy bit of the branch of the child.
 $unmark(val, child)$ sets to zero both the coalescing and the occupancy bits relative to the branch of the child.
  $is\_coal(val, child)$ returns {\em true} if the coalescing bit relative to the child is set to one.
  $is\_occ\_buddy(val, child)$ returns true if the occupancy bit relative to the buddy associated with the child
  is set to one.
  $is\_coal\_buddy(val, child)$ returns true if the coalescing bit relative to the buddy associated with the child
  is set to one.
In order to capture whether the child is the right or the left one, all these functions use a two-modulus operation applied to the index of the child.

We additionally use the following status-bit management function to detect whether a node, whose status bits are embedded within the $val$ bit-mask, is currently free:
\begin{description}
	\item[$is\_free(val)$ = $\neg (val~\&~{\tt BUSY})$]
\end{description}
This condition is verified when the node itself has not been reserved for some allocation operation (it is not occupied), and none of its sub-trees (left and/or right) has nodes currently reserved for allocations---none of the sub-trees is partially or fully occupied.

Similarly to most common allocators, our non-blocking buddy system exposes two API functions for either requesting a chunk of memory of (at least) a given size or for releasing some previously allocated memory chunk identified by its address. Both ${\tt tree}$[] and ${\tt index}$[] are initialized to zero at start-up. Recall that 0 does not correspond to any node of the tree since the initial element
of ${\tt tree}$[] is associated with index equal to 1.
Setting the entries of ${\tt index}$[] to zero indicates that none of the possible memory chunks (and of the corresponding addresses) managed by the buddy system at any level has been delivered for usage.

\subsection{Memory Allocation Algorithm}

The non-blocking memory allocation operation is divided in two algorithms, {\sc NBAlloc}() and {\sc TryAlloc}().
The pseudo-code of {\sc NBAlloc}(), which represents the memory allocation API actually exposed to the user, is reported in Algorithm \ref{nballoc}.
{\sc NBAlloc}() takes as input the size of the memory allocation request and, if such amount of memory is available as a set of buddies, it returns the address of a memory chunk big enough to fit the request.
If the size exceeds the overall memory allocatable by a single invocation, the allocation fails.
Differently, if it is smaller than the minimum amount managed within the buddy system, it is rounded to the allocation unit, namely the size associated with the leaves.
In any case, if the request size is legitimate, the target level of nodes to be considered for allocation is obtained by Rule \ref{eq:getlevel}
(line A5).
Once identified the right level, thus the range of indexes of nodes suitable for the allocation, these nodes are scanned in order to search for a free one.
Note that not necessarily such a search has to start from the first node at that level.
Rather, starting from scattered points will more likely lead concurrent allocations (bound to that same level) to target different free nodes, if any.
When a free node
is found (line A12), the allocation operation tries to ``reserve'' it by invoking the {\sc TryAlloc}() procedure.
This returns zero upon success.
Otherwise, it returns the index of the node that makes the allocation fail.
In the latter case, the algorithm moves to the next candidate node by exploiting the index returned by {\sc TryAlloc}() (lines A18-A19) to skip the whole sub-tree relative to the node causing the failure.
If no node at that level is found to be free, then the allocation operation fails, indicating that the current usage state of the buddy system is not compliant with the issued request.
%

When
the {\sc TryAlloc}() procedure succeeds at reserving a node, the relative address
with respect to ${\tt base\_address}$ is computed, and the corresponding entry of the ${\tt index}$[] array is updated to store the index of the reserved node (lines A15-A16).
Then,  such memory address is returned, indicating a successful allocation.

\begin{algorithm}[t]
\algosizefinal
\caption{Allocation - Part A}\label{nballoc}
	\algrenewcommand{\alglinenumber}[1]{\algprintlinenumber{ \makebox[20pt][r]{A#1:}}}
	\begin{algorithmic}[1]
		\Procedure{nballoc}{size\_t $size$}
		\Return void *
			\If {$size$ $>$ ${\tt max\_size}$}
				\State \Return false
			\EndIf
			\State $level \leftarrow $ $\lfloor$\Call{$log_2$} {$\frac{total\_memory}{size}$} $\rfloor$
			\If {$level > {\tt depth}$}
				\State $level \leftarrow {\tt depth}$	
			\EndIf		
			\State $starting\_from \leftarrow $ $2^{level-1}$
			\State $until\_to \leftarrow 2^{level}-1$
			\For{$i \leftarrow start\_from$ {\bf to} $until\_to$}
				\If {$is\_free({\tt tree}$[$i$]$)$}
					\State $failed\_at \leftarrow$ \Call{TryAlloc} {$i$}
					\If {$\neg failed\_at$}
						\State ${\tt index}[\frac{{\tt starting_i} - {\tt base\_address}}{{\tt min\_size}}] \leftarrow i$								
						\State \Return ${\tt starting_i}$
					\Else
						\State $d \leftarrow (1<<({\tt level_i}-{\tt level_{failed\_at}}))$
						\State $i \leftarrow (failed\_at+1)*d$
					\EndIf				
				\EndIf
			\EndFor \State{}
			\Return NULL
		\EndProcedure
	\end{algorithmic}
\end{algorithm}

\begin{algorithm}[t]
\algosizefinal
\caption{Allocation - Part B}\label{tryalloc}
	\algrenewcommand{\alglinenumber}[1]{\algprintlinenumber{ \makebox[20pt][r]{T#1:}}}
	\begin{algorithmic}[1]
		\Procedure{tryAlloc}{index $n$}
			\Return index
			\If {$\neg$ \Call{CAS}{$\&{\tt tree}[n]$, $ 0$, $ {\tt BUSY}$} \label{alg:tryalloc:occupynode}}
				\State \Return $n$
			\EndIf
			\State $current \leftarrow n$
			\While {${\tt level_{current}}$ $>$ ${\tt max\_level}$ \label{alg:tryallocupperbound}}
				\State $child \leftarrow current$
				\State $current \leftarrow current >> 1$
				\Do
					\State $curr\_val \leftarrow {\tt tree}[current]$
					\If{$curr\_val$ $\&$ ${\tt OCC}$ \label{alg:tryalloc:ifoccreturn}}
						\State \Call{FreeNode}{$n$, level($child$)}
						\State \Return $current$
					\EndIf
					\State $new\_val \leftarrow clean\_coal(curr\_val,child)$
					\State $new\_val \leftarrow mark(new\_val,child)$					
				\doWhile{ $\neg$ {\sc CAS}($\&{\tt tree}[current]$, $curr\_val$, $new\_val$)}
			\EndWhile
			\State \Return 0
		\EndProcedure
	\end{algorithmic}
\end{algorithm}

\begin{figure}[!t]
\centering
\includegraphics[width=\linewidth]{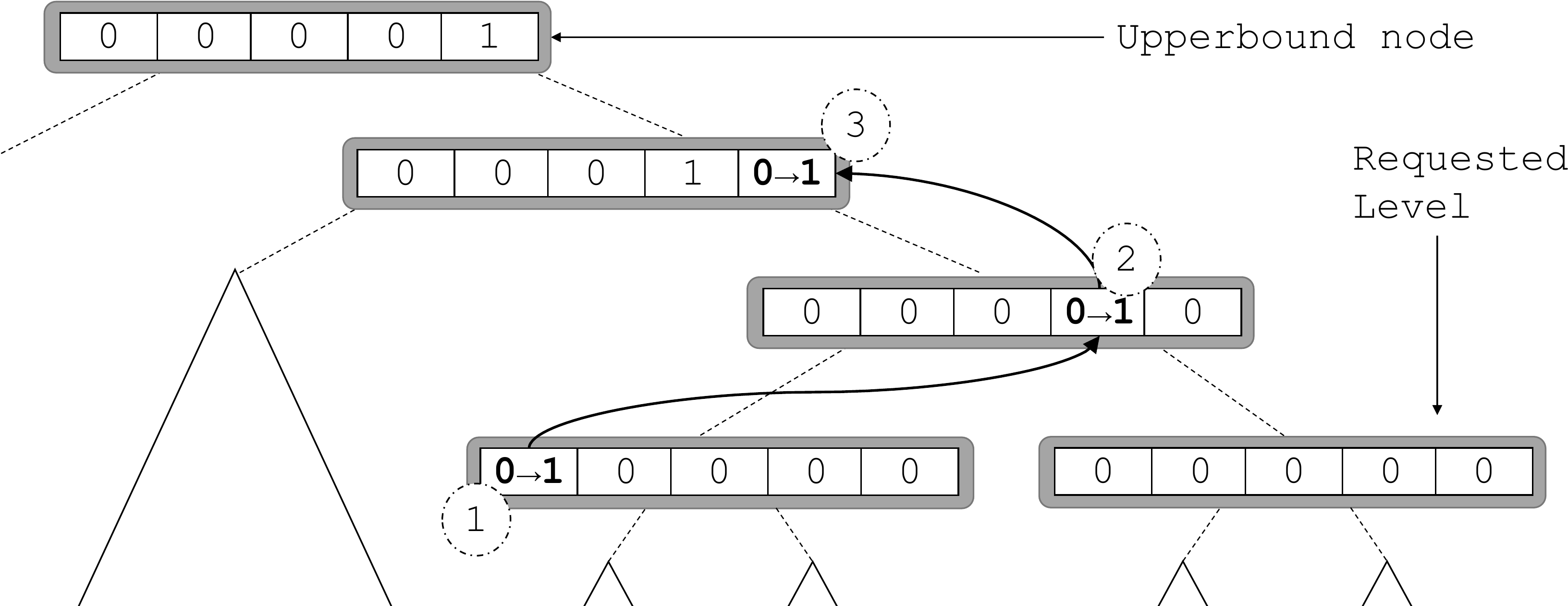}
\caption{Visual representation of {\sc TryAlloc} operations.}
\label{fig:steps1}
\end{figure}

The pseudo-code of {\sc TryAlloc}() is reported in Algorithm \ref{tryalloc} and a visual representation of its steps is shown in Figure \ref{fig:steps1}.
It takes as input the index of a node (previously observed as free) and tries to (i) occupy this node, and (ii) propagate the information about the occupancy up to the ancestor node belonging to the {\tt max\_level}.
The former task is carried out by relying on the CAS machine instruction (see line T2 corresponding to step 1 in Figure \ref{fig:steps1}), which tries to update all the occupancy bits in the status bit-mask of that node, by atomically checking if the status bits are still set to zero.
If the CAS fails, it means that something has concurrently changed on the state of the buddy system, affecting the ongoing operation.
Therefore, the memory allocation operation needs to be aborted and retried on a different node just as in the spirit of non-blocking coordination algorithms. 
Conversely, if the CAS instruction succeeds, the procedure continues by traversing the nodes along the path from the one that has been currently occupied towards the {\tt max\_level}.
This traversal is required to update the occupancy bits of the ancestor nodes so as to reflect that sub-tree has become partially occupied.
In this way the corresponding memory chunk will figure out as fragmented into lower level buddies, one of which is occupied, preventing other allocations from occupying an ancestor (a higher-level node).
For each node along the path towards the {\tt max\_level}, the procedure tries to mark its state as left or right occupied, depending on what branch we are backward traversing (e.g. step 2 and 3 in Figure \ref{fig:steps1}).

This operation is still performed via CAS machine instructions (see line T17) with the peculiarity that, if it fails, it can be retried since the update we are trying to perform can be still coherent with respect to the memory allocation operation we are carrying out and other concurrent operations.
As an example, a CAS on the node we are traversing may fail since a concurrent operation updates the occupancy bit associated with the other branch of the tree, or even the same branch.
The only scenario where the failure of the CAS at line T17 indicates that the currently carried out memory operation needs to be aborted
is when another operation updated the (fully) ${\tt occupied}$ bit.
In this case, another operation has exactly reserved that node---and the corresponding memory chunk---for a concurrent (or already-finalized) memory allocation.
Hence, we cannot fragment that chunk to reserve some chunk at lower levels.
We also note that, while attempting to left/right occupy a node along the traversal, the corresponding left/right coalescing bit is set to the value 0.
As we will clarify while explaining memory release operations, this needs to be done in order to make conflicting releases
be aware that the branch is again involved in a new allocation and then cannot be marked as free.

If the max-level node is reached and updated along the backward traversal, the node originally targeted when starting {\sc TryAlloc}() can be considered as correctly taken.
Conversely, if some node with the ${\tt occupied}$ bit set is found along the path, the alloc fails on the current memory block and nodes updated along the backward traversal have to be
reverted.
This is done by invoking the {\sc FreeNode}() procedure  in Algorithm \ref{free}, which is also used to support non-blocking memory release operations in our buddy system.
In such a scenario, as aforementioned, the {\sc TryAlloc}() procedure ends returning the index of the node for which the allocation has failed.

\subsection{Memory Release Algorithm}

A memory release operation is composed by three phases.
In the first phase, the ancestors of the node to be released are marked as coalescing, in order to notify that a free operation is in place along the corresponding path of the tree. 
In the next phase, the node to be released is marked as free by resetting all its occupancy bits.
During the last phase, all the nodes previously marked as coalescing are updated again to notify that the sub-tree involving the just released node is actually free---therefore it can serve again
memory requests.
As hinted before, this last update may conflict with a concurrent allocation operation and fail thanks to the CAS semantic---left/right occupancy bits remain set since the memory has been already reused.
The first two steps are implemented by the {\sc FreeNode}() procedure (see Algorithm \ref{free}), while the last one is carried out by the {\sc Unmark}() procedure (see Algorithm \ref{unmark}).

The {\sc FreeNode}() procedure is not directly exposed to the user.
It is instead encapsulated by the {\sc NBFree}() procedure, which is the actual memory release API in our buddy system.
This procedure receives the memory address corresponding to the chunk to be released, computes the relative index of the corresponding node by inspecting the ${\tt index}$[] array, and triggers the execution of the {\sc FreeNode}() procedure.

{\sc FreeNode}() takes as input the index of the node to be released and an {\em upper-bound}, which identifies the upper-level to be reached along the backward traversal associated with its execution.
If the {\sc FreeNode}() procedure is invoked by {\sc NBFree}(), the upper-bound is set to {\tt max\_level}, making it traverse the tree up to the level corresponding to the maximum allocatable size.
In this case we are releasing a previously allocated node and, thus, the status bits need to be reflected up to the maximum useful level of the tree.
Differently, if the procedure is invoked by a failed {\sc TryAlloc}() (see line T12), the upper-bound is set to the level of the last node updated by a {\sc TryAlloc}(), during an aborted memory allocation.
As discussed before, this execution path is related to the need for {\sc TryAlloc}() to restore the status bits of nodes that where involved in the traversal towards the maximum level, which was interrupted because of the discovery of an already occupied node.

\begin{figure}[t]
\centering
\includegraphics[width=\linewidth]{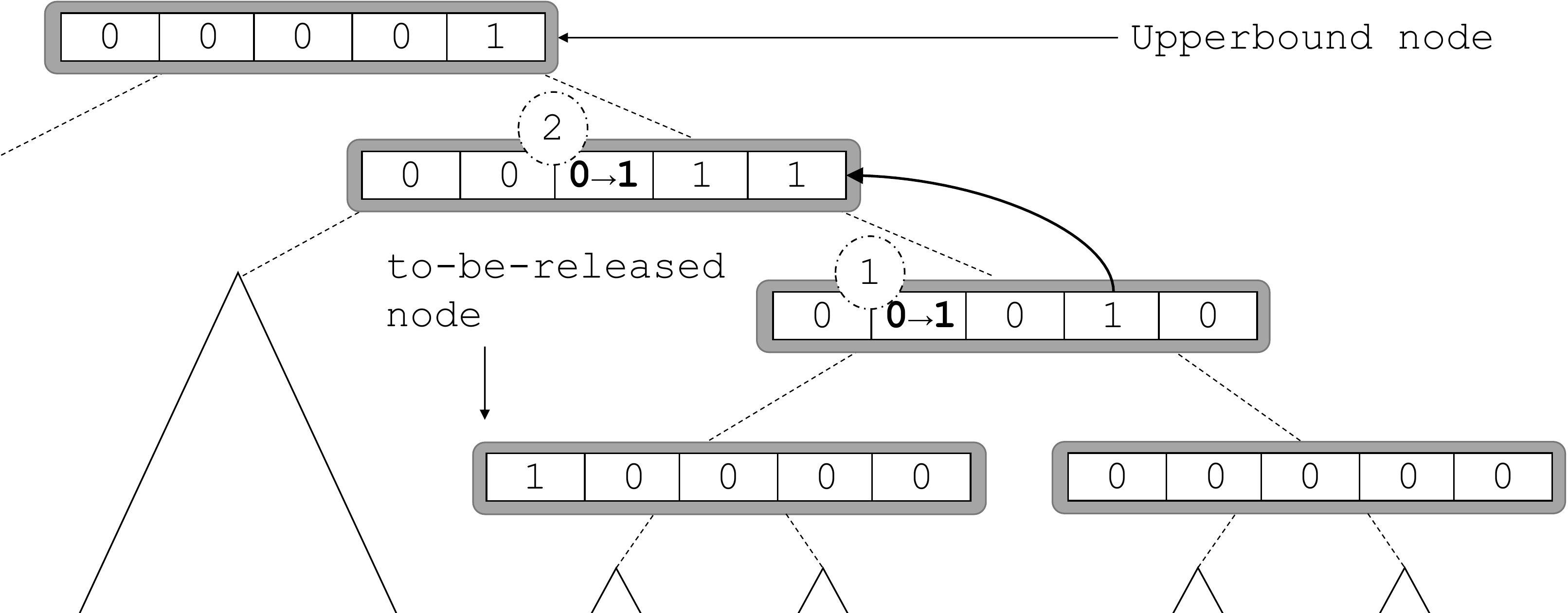}
\caption{First phase of the {\sc FreeNode} (lines F2-F18).}
\label{fig:steps2}
\end{figure}

While traversing all the nodes up to the upper-bound, {\sc FreeNode}()
atomically sets the coalescing bit of the correct position, via CAS instruction (see line F10 in Algorithm \ref{free}).
A visualization of this phase is shown in Figure \ref{fig:steps2}.

\begin{figure}[!t]
\centering
\includegraphics[width=\linewidth]{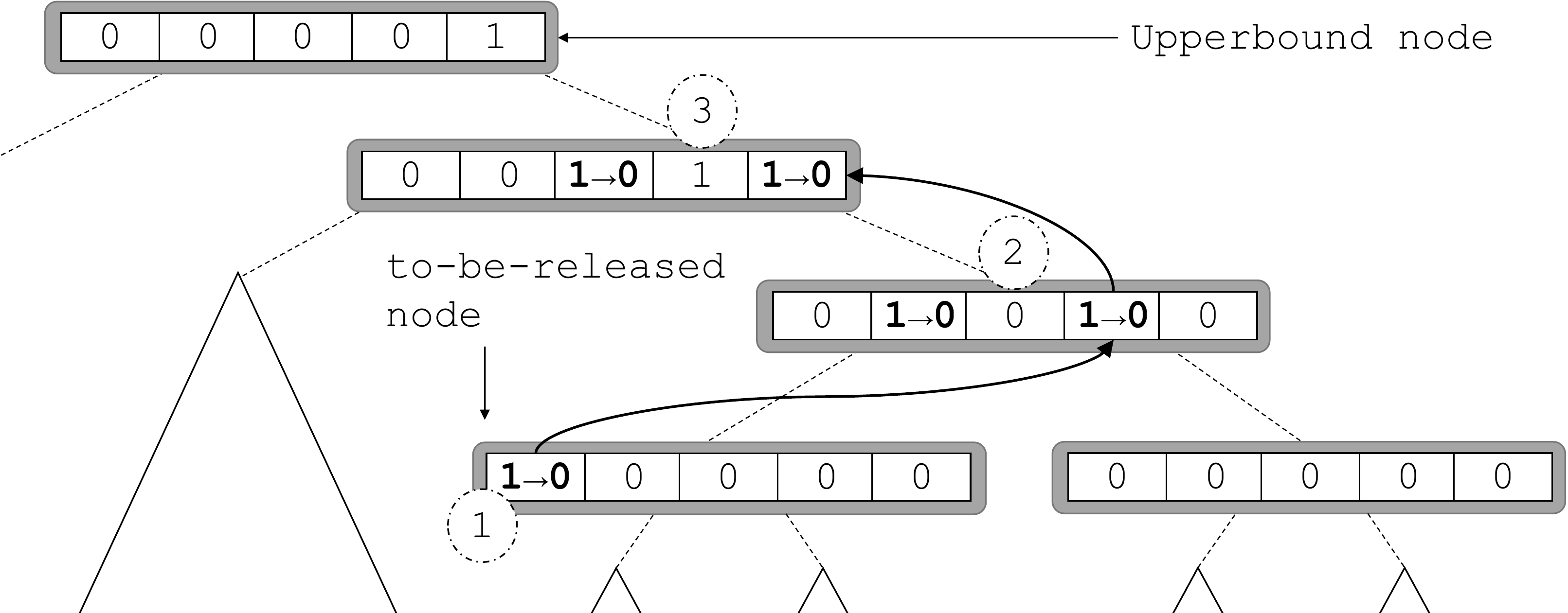}
\caption{Second (line F19) and third ({\sc Unmark}) phase of the {\sc FreeNode}.}
\label{fig:steps3}
\end{figure}

If along this path a buddy is detected as occupied by other allocations, the climb is early arrested, since the corresponding sub-tree cannot be considered free---this is the case of Figure \ref{fig:steps2}, where the right subtree of the node below the upper-bound is already fragmented.
Once the first phase is concluded, {\sc FreeNode} can start to signaling that the interested node has been released, whose steps are sketched in Figure \ref{fig:steps3}.
In particular, the node to be released can be updated by resetting its occupancy bits.
This takes place by simply writing zero on its status bits (see line F19 in Algorithm \ref{free} corresponding to step 1 of Figure \ref{fig:steps3}).

\begin{algorithm}[t]
\algosizefinal
\caption{Memory Release}\label{free}
\algrenewcommand{\alglinenumber}[1]{\algprintlinenumber{ \makebox[20pt][r]{F#1:}}}
	\begin{algorithmic}[1]
		\Procedure{NBFree}{void *$addr$}
			\State{$n$ $\leftarrow${\tt index}$[\frac{addr - {\tt base\_address}}{{\tt min\_size}}]$}		
			\State{\Call{FreeNode}{$n$,${\tt max\_level}$}}
		\EndProcedure
	\end{algorithmic}

	\begin{algorithmic}[1]
		\Procedure{FreeNode}{index $n$, index $upper\_bound$}
		\State{$current \leftarrow n >> 1$}
		\State{$runner \leftarrow n$}
		\While{${\tt level_{runner}}$ $>$ $upper\_bound$}
			\State{ $or\_val \leftarrow {\tt COAL\_LEFT}>>(mod_2(current))$ \label{alg:freenode:setascoalesce}}
			\Do
				\State $cur\_val \leftarrow {\tt tree}[current]$
				\State $new\_val \leftarrow cur\_val$  $|$ $or\_val$
				\State $old\_val$ $\leftarrow$\\ \hfill \Call{CAS}{$\&{\tt tree}[current]$,$cur\_val$,$new\_val$}
			\doWhile {$old\_val$ $\neq$ $cur\_val$}
			\If{$is\_occ\_buddy(old\_val,runner)$ $\wedge$\\ \hfill $\neg is\_coal\_buddy(old\_val,runner)$}
				\State \textbf{break}
			\EndIf
			\State{$runner \leftarrow actual$}
			\State{$current \leftarrow current >> 1$}
		\EndWhile
		\State {${\tt tree}[n] \leftarrow 0$ \label{alg:free:releasenode}}
		\If{$n \neq upper\_bound$}
			\State \Call{Unmark}{$n$,$upper\_bound$}
    	\EndIf
		\EndProcedure
	\end{algorithmic}
\end{algorithm}

\begin{algorithm}[!t]
\algosizefinal
\caption{Unmark}\label{unmark}
\algrenewcommand{\alglinenumber}[1]{\algprintlinenumber{ \makebox[20pt][r]{U#1:}}}
	\begin{algorithmic}[1]
		\Procedure{Unmark}{index $n$, index $upper\_bound$}
		\State{$current \leftarrow n$}
		\Do
			\State {$child \leftarrow current$}
			\State{$current \leftarrow current >> 1$}
			\Do	
				\State $curr\_val \leftarrow {\tt tree}[current]$
				\If {$\neg is\_coal(curr\_val, child)$ \label{alg:unmark:stopfaster}}	
					\State \Return
				\EndIf
				\State $new\_val \leftarrow unmark(curr\_val,child)$
			\doWhile { $\neg$ CAS($\&{\tt tree}[current]$,$ curr\_val$, $ new\_val$)}
		\doWhile { (${\tt level_{current}}$ $>$ $upper\_bound$) $\wedge$\\ \hfill $\neg is\_occ\_buddy(new\_val,child) $}
		\EndProcedure
	\end{algorithmic}
\end{algorithm}


The last phase is responsible of propagating the release of the interested node up to the upper-bound and, possibly, merging buddies.
This is achieved by invoking {\sc Unmark}(),  which traverses the nodes, from the one to be released, towards the upper-bound, cleaning the coalescing and the occupancy bits of the traversed nodes (see steps 2 and 3 of Figure \ref{fig:steps3}).
For each node met, the first step is to verify whether the coalescing bit is still set: if it is not, the procedure returns, finishing the release operation.
As hinted before, this scenario is possible if some allocation\slash release in the same sub-tree of the target node has already occupied/released that coalescing sub-tree.

Conversely, if the coalescing bit is still set, the procedure tries to clean both the coalescing and the occupancy bits atomically.
 This operation is done via a CAS in a retry-cycle in order to manage the case in which the coalescing bit has been reset by some concurrent operation, meaning that the resource has already been reused/released.
If during this procedure some nodes in the same sub-tree have been allocated, the relative occupancy bits have not to be reset.
Then, the procedure checks if the buddy occupancy bit is set at each step and, in the positive case, it returns.
In fact, similarly to the first phase, if such buddy is occupied, we cannot propagate node releasing and, thus, merge buddies up to the higher level, since the chunks associated with higher level nodes are still fragmented.


Overall, beyond providing non-blocking capabilities while allocating or releasing memory at a given level, our buddy system allows fragmenting and merging operations---which logically move nodes across different levels within the allocation scheme---still in non-blocking fashion. This operation, that usually has to be explicitly performed, here is carried out automatically while performing allocations/releases.

\subsection{4-Levels Optimization}
\label{4lvl}
In order to execute an atomic RMW instruction, the CPU core takes an exclusive access to the relative memory cache line, delaying its access to other cores.
Reducing the number of RMW instructions executed along the critical path of any thread in a non-blocking algorithm is therefore an important aspect in terms of further performance improvements.


In the previously presented solution, a thread that performs a memory allocation or a memory release has to execute a number of RMW instructions equal to at least the depth of the node involved in the allocation/release respect the ${\tt max\_level}$ within the tree. However, since the number of status bits required to represent the state of each node is significantly smaller than the word size---which is 64-bits on nowaday's conventional machines---is possible to reduce such amount of RMW instructions by packing a bunch, namely a node and some levels descendant from it, in the same status bit-mask. With this organization a thread is able to update at one time the state of multiple levels with only one RMW atomic instruction.
On the other hand, the coherence with the original specification is guaranteed by the fact that all state-update operations are performed via CAS, which succeeds only if the status bit-mask has not changed in the meanwhile.

\begin{figure}[t]
\centering
\includegraphics[width=\linewidth]{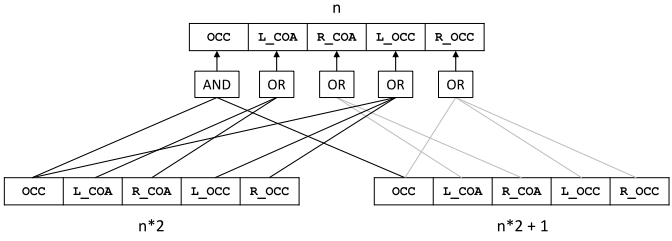}
\vspace*{-0.35cm}
\caption{Derivation of node state by children' state.}
\label{fig:log}
\end{figure}

Another important consideration is that, given a generic node in the tree, its state can be derived by looking at the state of its children, as shown in Figure \ref{fig:log}.
In fact, the partial occupancy of a node---say its left/right occupancy---can be computed with a logical OR operation on the (partial and full) occupancy bits of the children nodes. In the same way, the coalescing bits can be derived by the ones of the children.

Moreover, when a node is actually set as ${\tt occupied}$ also its children can be logically considered to reside in the same state, since they cannot be seen as individual fragments of memory some of which considerable as freely available for allocation.
In the same way, occupying two buddies looks to be the same as fully occupying the parent node.
This means that, similarly to the partial occupancy, the fully occupancy of a node can be computed with a logical AND operation on the ${\tt occupied}$ bits of its children.

\begin{figure}[t]
\centering
\includegraphics[width=0.7\linewidth]{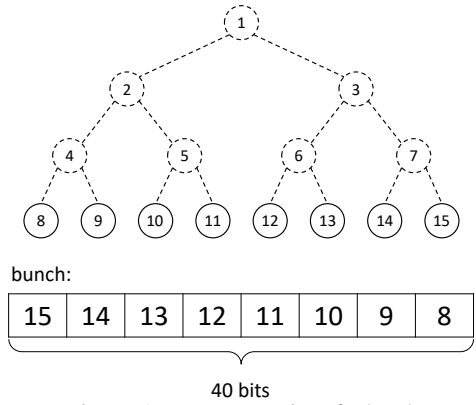}
\vspace*{-0.35cm}
\caption{Array representation of a bunch.}
\label{fig:bunch}
\end{figure}

Such reasoning can be recursively applied to all the ancestors of a node, hence a sub-tree starting from a given node can be represented by the nodes in its lower level.
Then, considering a word-size equal to $w$ bits, and a status bit-mask of size $s$ bits, we can pack in a single variable a bunch of depth equal to $d$, with $2^{d}*s < w$, representing $2^{d+1}-1$ nodes of the original tree. In our case, we are able to manage 4 levels, namely 15 nodes with 8 nodes in the lower level (40 bits), in a single 64-bit word, reducing the number of atomic RMW operations by a factor of 4. A representation of this transformation is shown in Figure \ref{fig:bunch}.

Of course, this means that, when an operation is performed on a node that is not in a lower level of a bunch, its state has to be decoupled from its descendant in the same bunch.
To practically implement the operation in this variant of our data structure, each node in the ${\tt tree}$[] array stores a pointer to its bunch and its position inside it (computed with the same rule used in the original solution).
The algorithms for the 4-level solution are pretty similar to those shown before, with two main differences.
First, the direct allocation\slash release of a node has to check and then set the state of all the nodes in the sub-tree in the lower level of the bunch. In particular, given a node $n$ with depth $d_n$ and its position inside the bunch $b_n$, the corresponding bunch-leaf nodes have position $p$ in the bunch equal to:	
\[ p \in  [b_n*2^{3-mod_4(d_n)},(b_n+1)*2^{3-mod_4(d_n)}-1]\]
Second, updates of the bunches up to the root have to be executed with a step of 4 levels in order to update each time the relative bunch-node.

\section{Experimental Results}
\label{data}

In our experimental evaluation, the compared allocators do not use pre-reserving, hence they are tested as actual back-end allocation services.
This is compliant with our objective, since we focus on improving the efficiency of a back-end memory manager that could be used in combination with any pre-reserving (and thread binding to pre-reserved memory) policy taken from the literature.
As already hinted, pre-reserving is an orthogonal technique to improve memory allocation performance with respect to making a back-end allocator more scalable, via non-blocking algorithms as in our approach.

We have compared the performance of our non-blocking buddy system with various alternatives: 1) the buddy allocator in \cite{altro-buddy} denoted as {\tt buddy-sl}, which is still based on a tree data structure, but synchronizes concurrent accesses using spin-locks,
2) the Linux buddy system, which is instead based on a multi-list data structure, and exploits spin-locks for handling concurrency.
Each back-end allocator has been set up in a single-instance configuration, since we are interested in evaluating the benefits/shortcomings of our approach when concurrent requests can actually insist on the same allocation data structure instance, rather
than exploiting data-separation via multi-instances.
In our tests we have assessed both the original and the 4-level optimized versions of our buddy system, which we denote as {\tt 1lvl-nb} and {\tt 4lvl-nb}, respectively. Also, we include data related to our own data structure with the variant that, rather than using RMW instructions to make it non-blocking, we synchronize the accesses in a blocking manner by using a unique (global) spin-lock. This configuration, named as  {\tt 1lvl-sl} and {\tt 4lvl-sl} for the two different levels' organizations, respectively, has been included just to study the effects of non-blocking operations against a wider set of blocking solutions, each based on a different implementation.

\begin{figure*}[h!]
\centering
\vspace*{0.5cm}
\includegraphics[width=0.30\linewidth, clip, trim = 3 2 6 5]{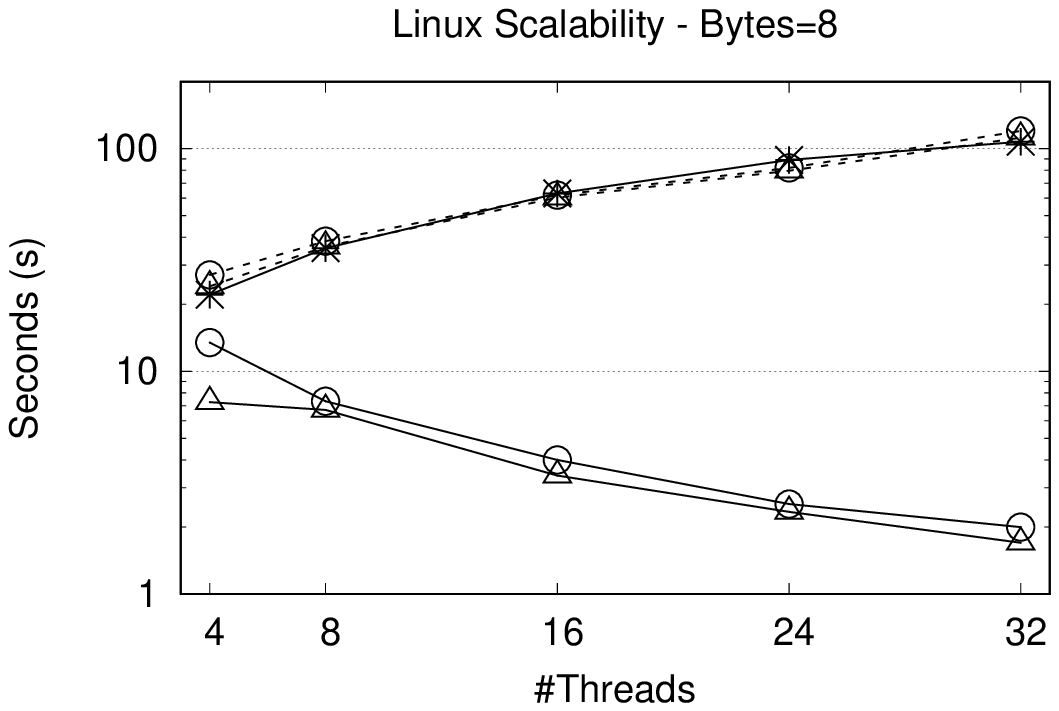}   		 \label{fig:tbls8}
\includegraphics[width=0.30\linewidth, clip, trim = 3 2 6 5]{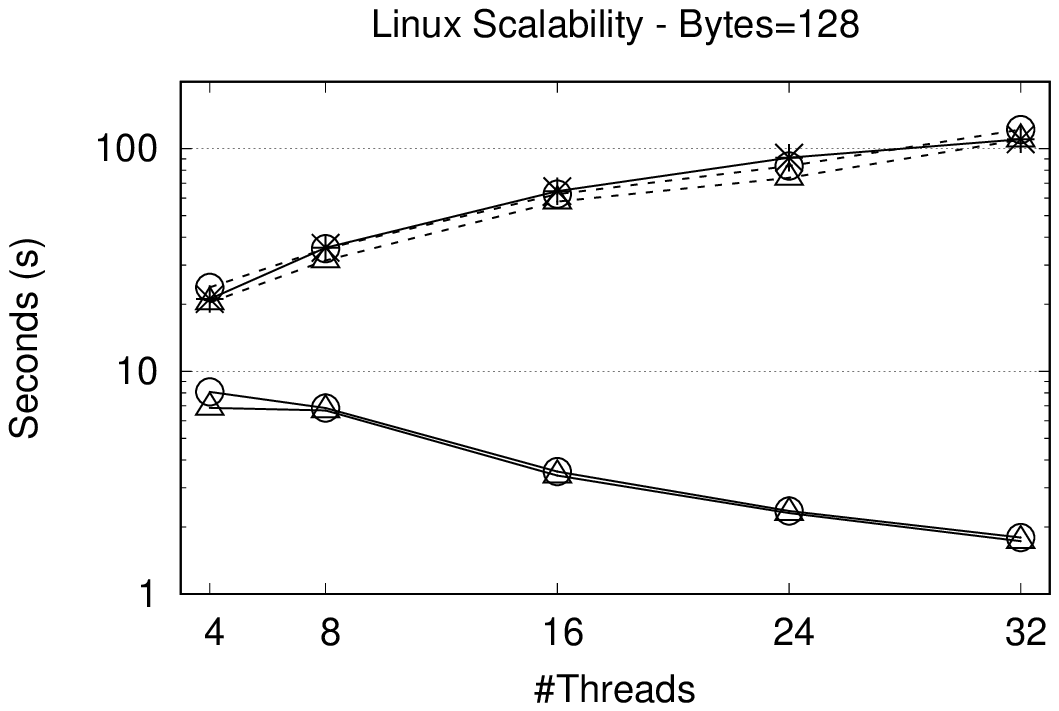}     \label{fig:tbls128}
\includegraphics[width=0.30\linewidth, clip, trim = 3 2 6 5]{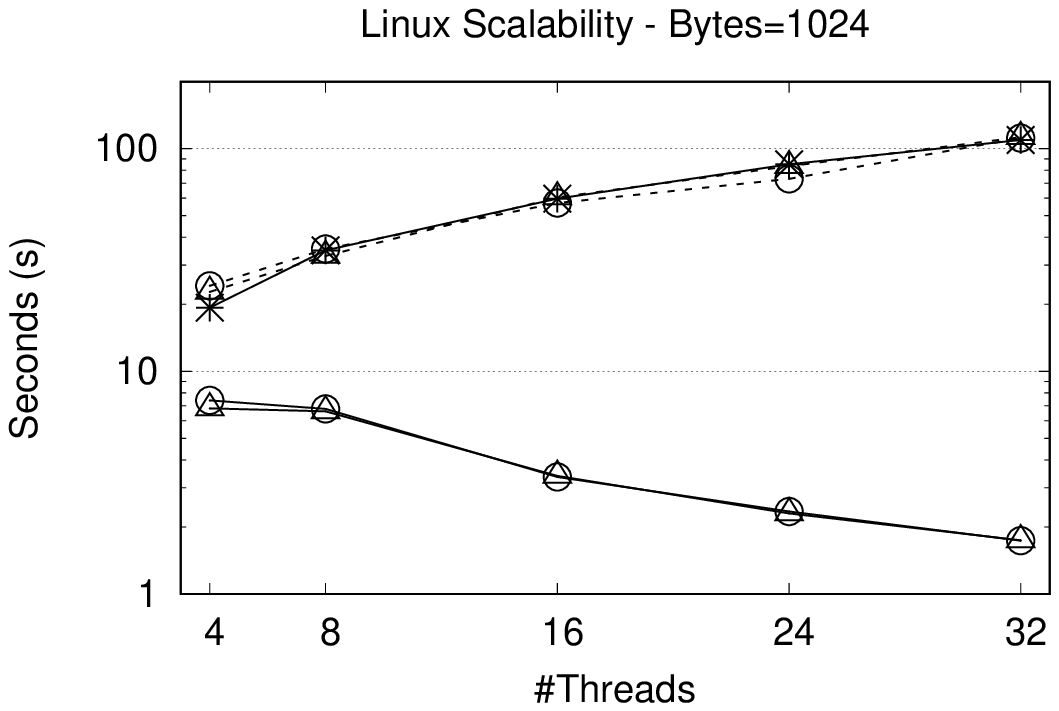}   \label{fig:tbls1024}
\includegraphics[width=0.8\linewidth, clip, trim = 10 0 5 210]{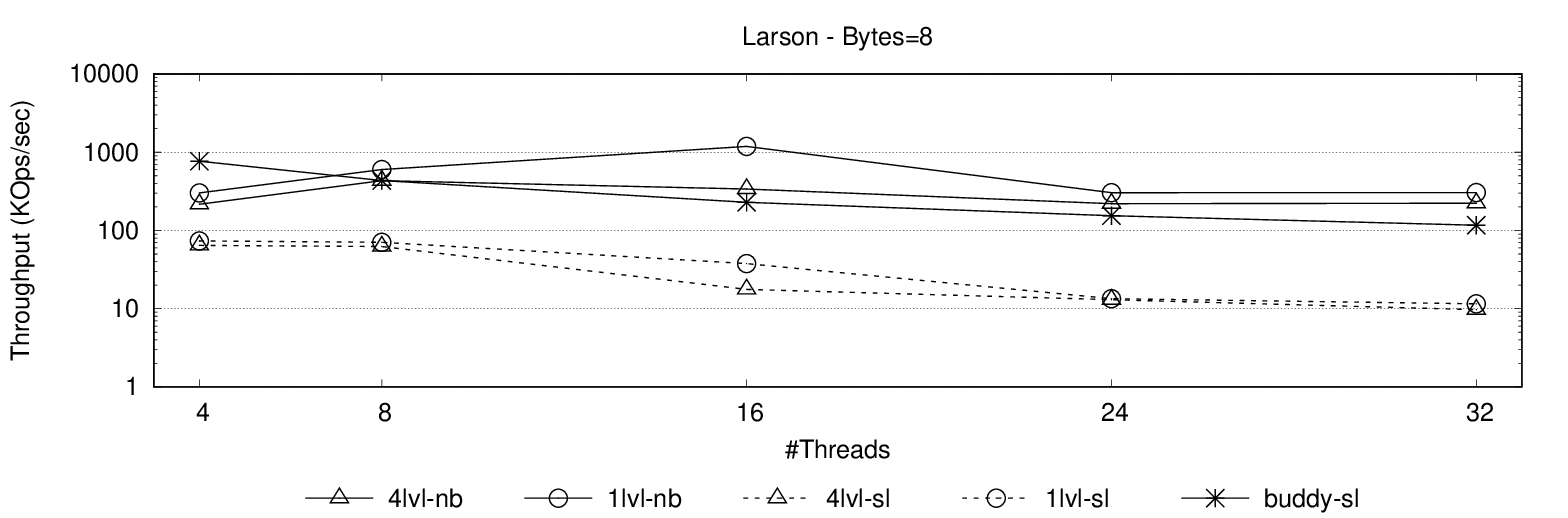}
\caption{Execution times - Linux Scalability benchmark.}
\label{fig:tbls}
\vspace*{0.5cm}
\includegraphics[width=0.30\linewidth, clip, trim = 3 2 6 5]{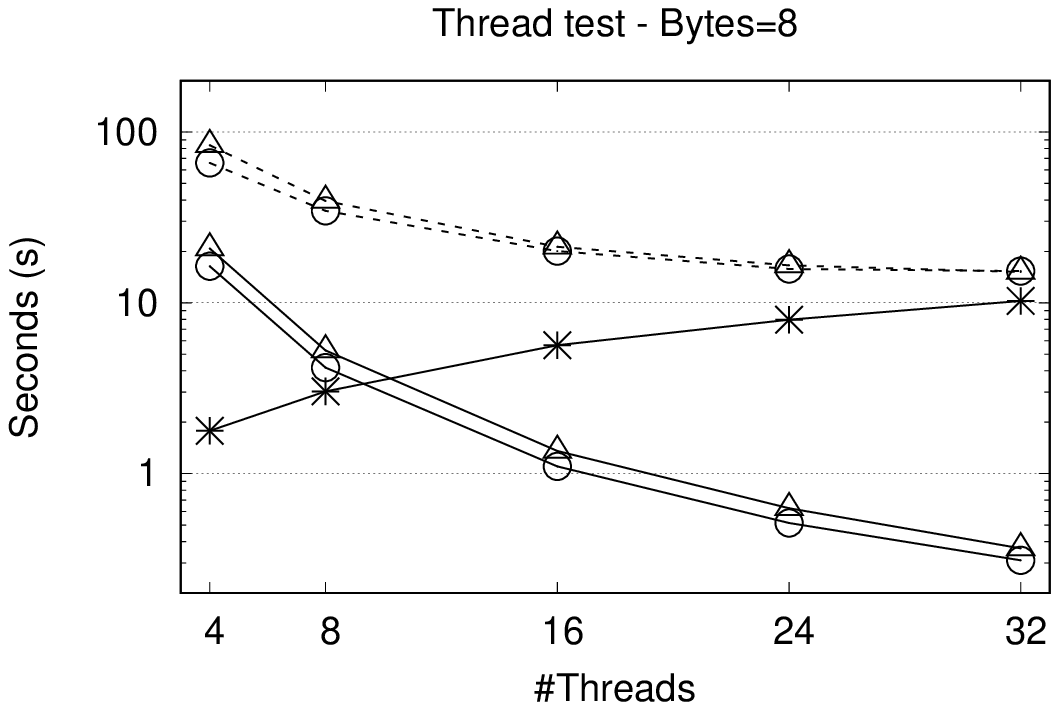}   \label{fig:tbtt8}
\includegraphics[width=0.30\linewidth, clip, trim = 3 2 6 5]{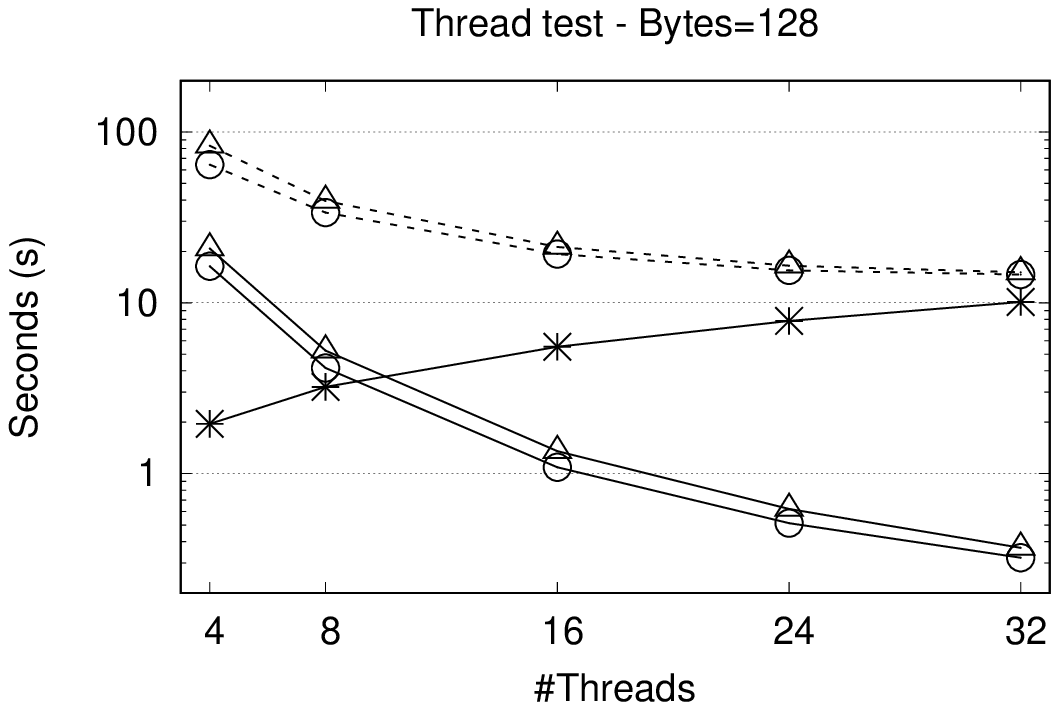}   \label{fig:tbtt128}
\includegraphics[width=0.30\linewidth, clip, trim = 3 2 6 5]{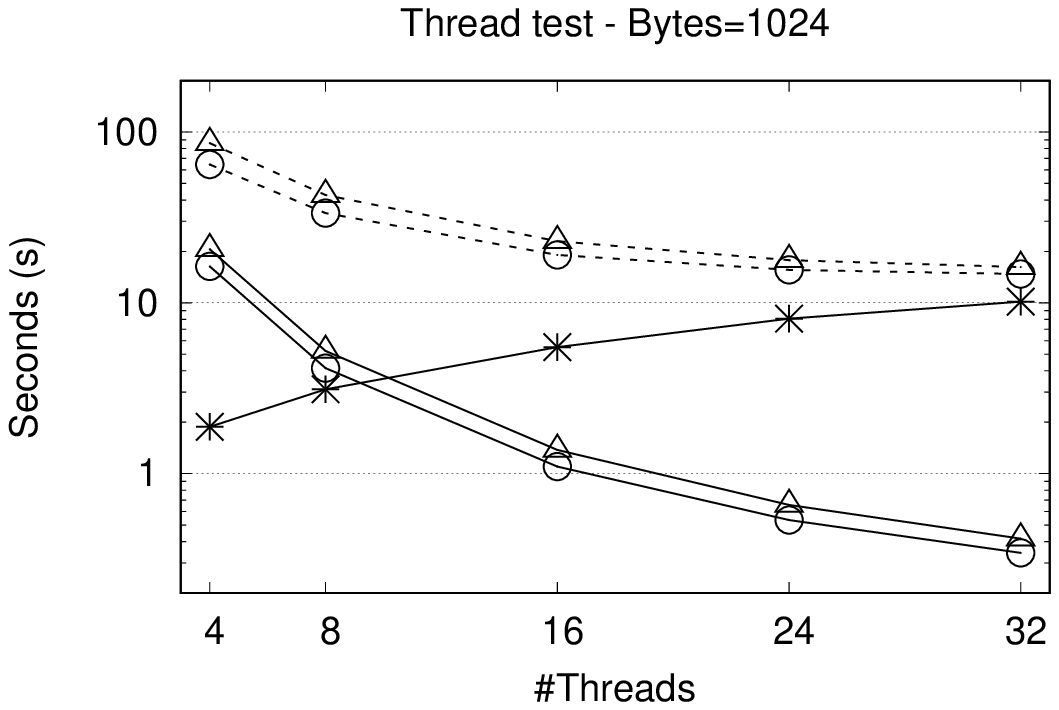}   \label{fig:tbtt1024}
\includegraphics[width=0.8\linewidth, clip, trim = 10 0 5 210]{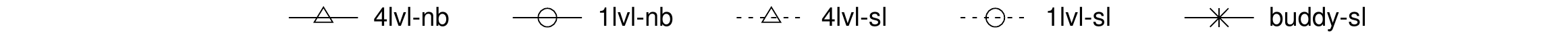}
\caption{Execution times - Thread Test benchmark.}
\label{fig:tbtt}
\vspace*{0.5cm}

\centering
\includegraphics[width=0.3\linewidth, clip, trim = 3 2 3 5]{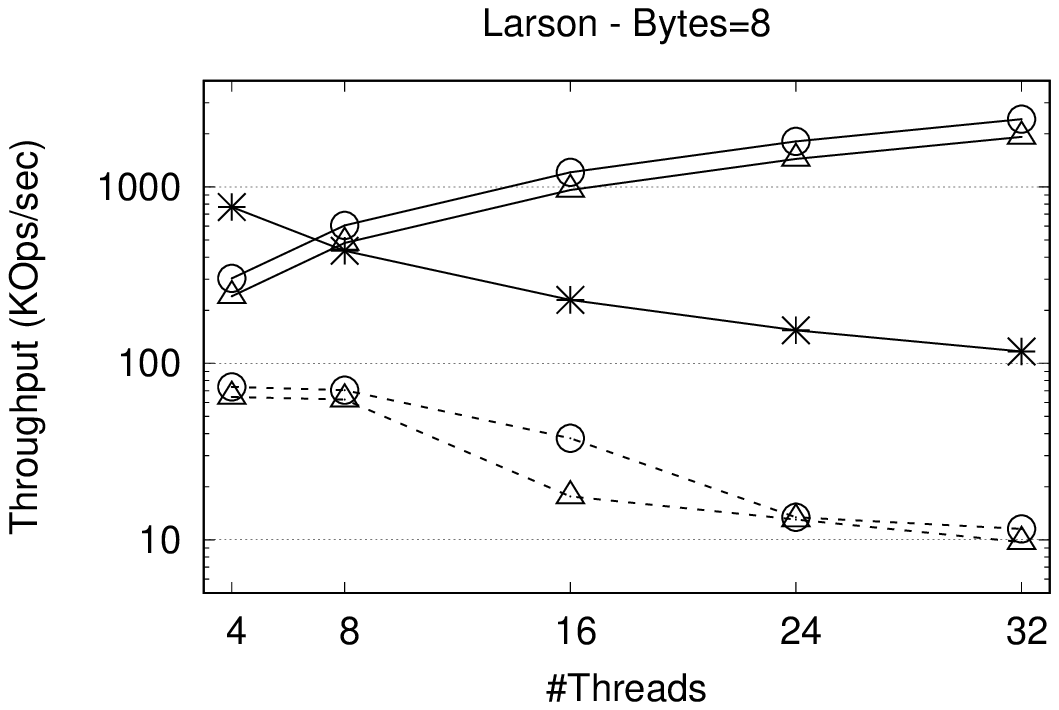}   \label{fig:lrsn8}
\includegraphics[width=0.3\linewidth, clip, trim = 3 2 3 5]{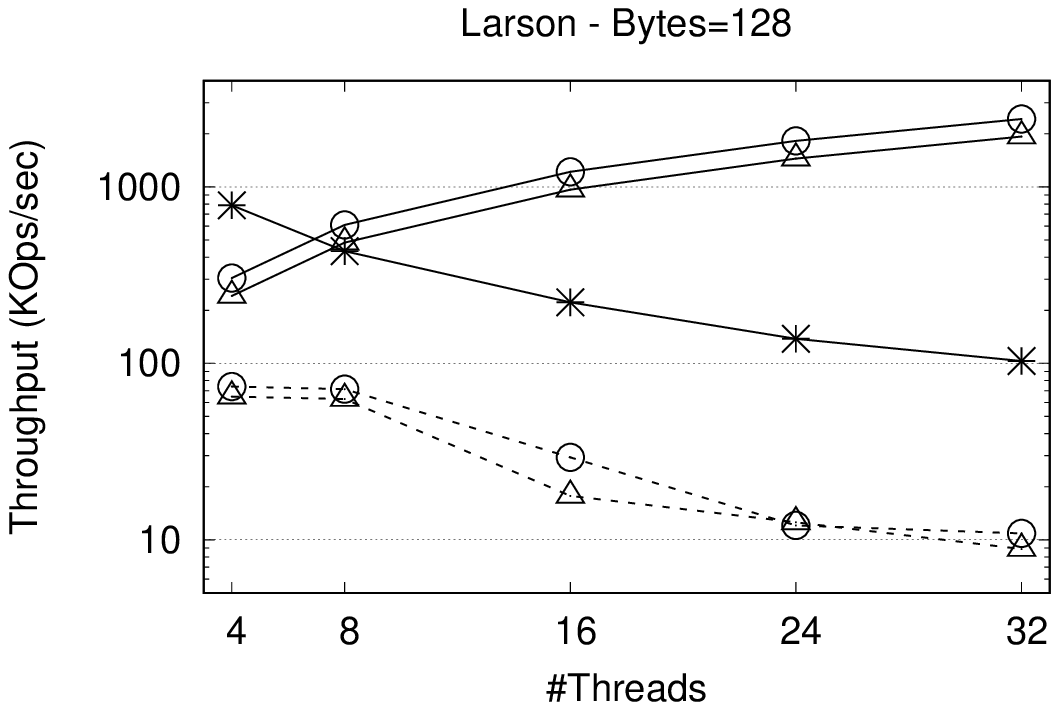}   \label{fig:lrsn128}
\includegraphics[width=0.3\linewidth, clip, trim = 3 2 14 5]{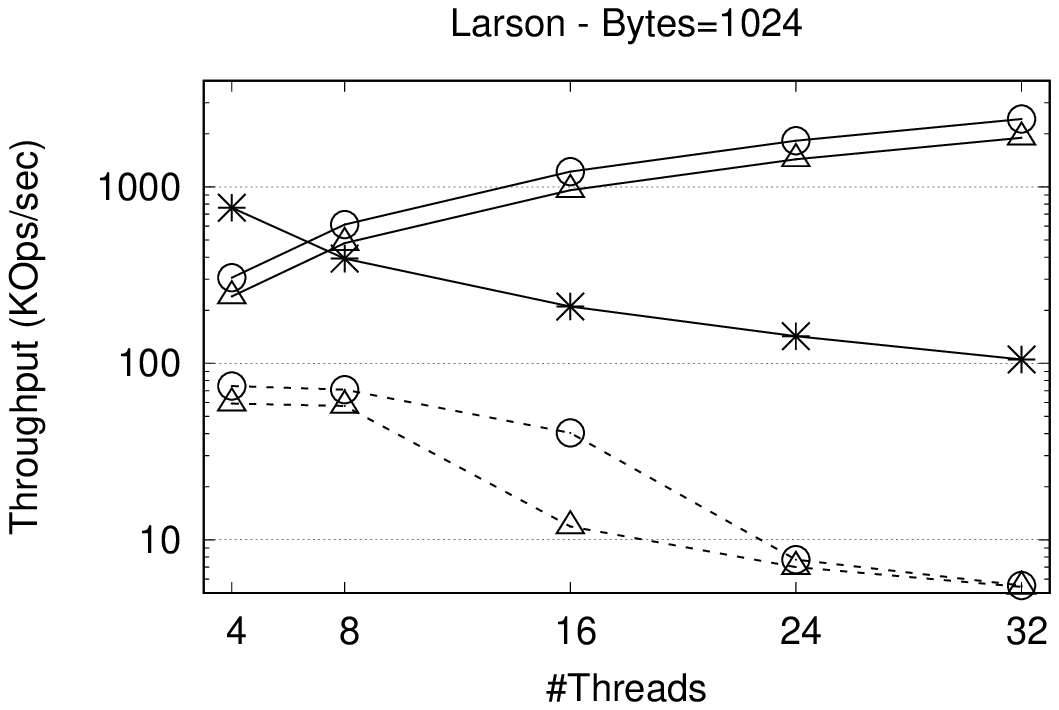}   \label{fig:lrsn1024}
\includegraphics[width=0.8\linewidth, clip, trim = 10 0 5 210]{./img/LEGEND}
\caption{Throughput - Larson benchmark.}
\label{fig:lrsn}
\vspace*{0.5cm}

\centering
\includegraphics[width=0.30\linewidth, clip, trim = 3 2 6 5]{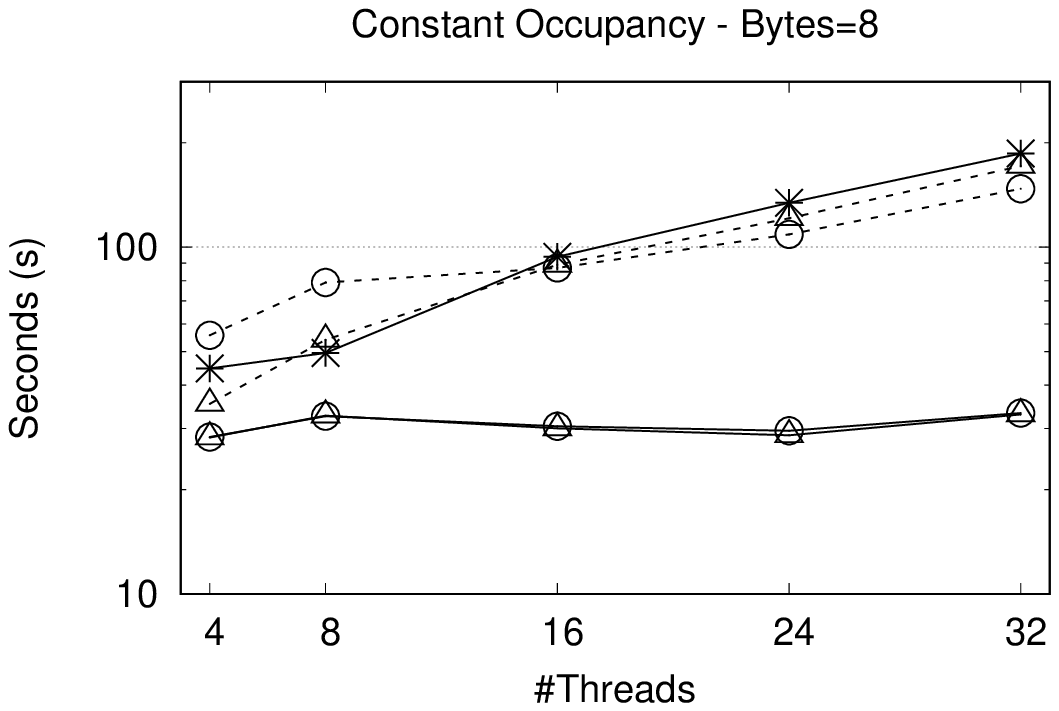}   \label{fig:tbfs8}
\includegraphics[width=0.30\linewidth, clip, trim = 3 2 6 5]{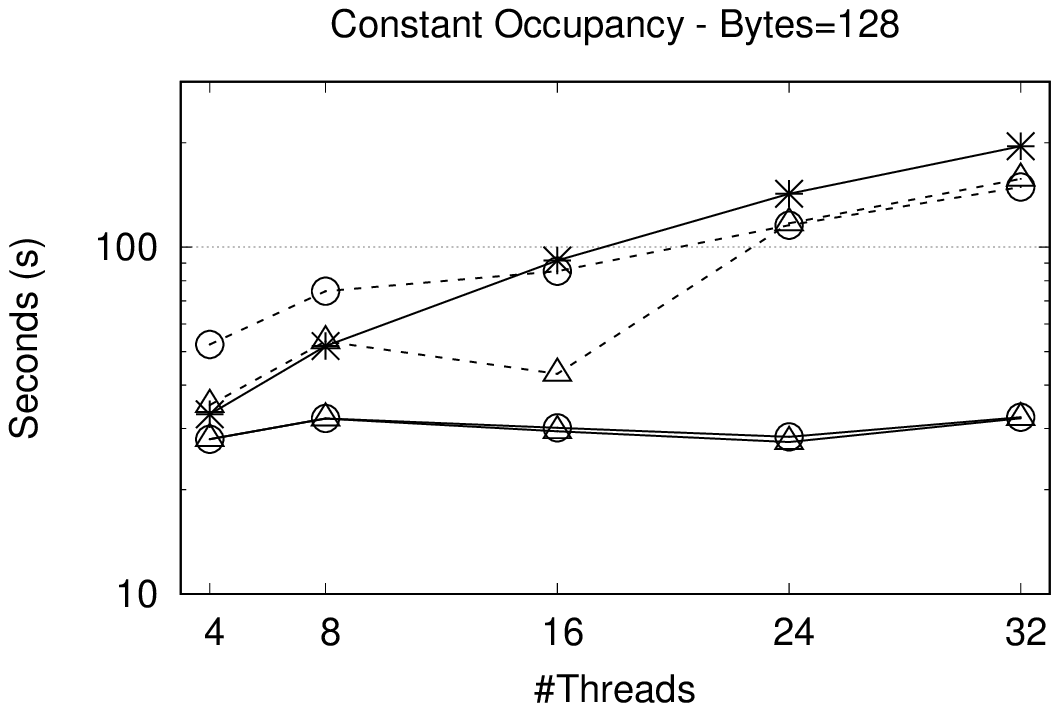}   \label{fig:tbfs128}
\includegraphics[width=0.30\linewidth, clip, trim = 3 2 14 5]{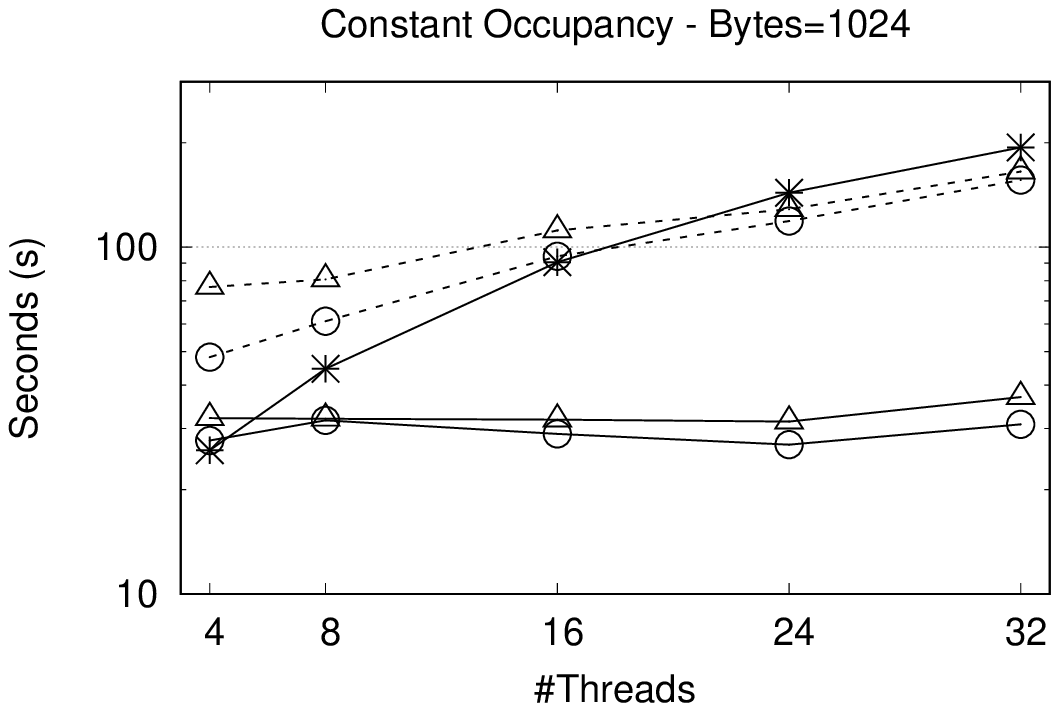}   \label{fig:tbfs1024}
\includegraphics[width=0.8\linewidth, clip, trim = 10 0 5 210]{./img/LEGEND}
\caption{Execution times - Constant Occupancy benchmark.}
\label{fig:tbfs}
\end{figure*}

\remove{

\begin{figure*}[h!]
\centering
\subfigure[]{\includegraphics[width=0.4\linewidth, clip, trim = 3 2 6 5]{{./data/plots/TBLS/TBLS-8}.eps}   		 \label{fig:tbls8}}
\centering
\subfigure[]{\includegraphics[width=0.4\linewidth, clip, trim = 3 2 6 5]{{./data/plots/TBTT/TBTT-8}.eps}   \label{fig:tbtt8}}
\centering
\subfigure[]{\includegraphics[width=0.4\linewidth, clip, trim = 3 2 6 5]{{./data/plots/TBFS/TBFS-8}.eps}   \label{fig:tbfs8}}
\centering
\subfigure[]{\includegraphics[width=0.4\linewidth, clip, trim = 3 2 3 5]{{./data/plots/LRSN/LRSN-8}.eps}   \label{fig:lrsn8}}
\includegraphics[width=\linewidth, clip, trim = 10 0 5 210]{./img/LEGEND}
\caption{Experimental evaluation of the user-space allocators with several benchmarks. The x-axis represents the thread count, while the y-axis represents the execution time (the lower the better) for figures (a-c) and the throughput (the higher the better) for figure (d).}
\label{fig:tbfs}
\end{figure*}

}
For the assessment we used 4 different test scenarios, 3 of which have been taken from the literature, and one has been specially devised for this study. The first is the Linux scalability test \cite{Lever2000},  where threads continuously execute an allocation/release pattern, with fixed size, for at least 20\,000\,000/{\tt num\_threads} times. The second is the Thread Test presented in \cite{berger}, where threads carry out 10\,000/{\tt num\_threads} allocations of a given size, and then release the acquired memory, executing this pattern in a cycle of at least 200 steps. The third is the one by Larson in \cite{Larson}, where the behavior of a Web server is simulated, by having a thread reserving memory for (emulated) operations, and then releasing it for usage by other operations.
According to its specification, this test is aimed at assessing the operations throughput over a time window of 10 seconds. The fourth---the one we propose---which we call Constant Occupancy, is based on having each thread initially allocating a pool of chunks of different sizes, with larger amount of allocations bound to smaller chunk sizes, and then it performs 20\,000\,000/{\tt num\_threads} deallocations/allocations by randomly selecting the element to be deallocated---hence the corresponding deallocation size---and using this same size for the subsequent allocation. Compared to the others, this test more prominently uses allocations/deallocations involving chunks of different size, and tends to keep constant the factor of occupancy of the buddy system.
All the tests have been carried out on a 64-bit NUMA HP ProLiant server equipped with four 2GHz AMD Opteron 6128 processors and 64 GB of RAM. Each processor has 8 cores, for a total of 32 CPU-cores. The used operating system is Linux, with kernel version 3.2.

\remove{
In a first set of experiments, we focuses on user space allocation, thus delaying the comparison with the Linux buddy system to a second part of the analysis. In these experiments we have configured the user-space allocators to manage chunks of minimal size equal to 8 bytes, and maximal size set to 16KB. Also, we used different allocation/deallocation sizes ranging from 8 bytes to 1024 bytes---for the Constant Occupancy test these sizes indicate the minimum ones among those managed, while the maximum ones are set to be 16 times larger. In Figures \ref{fig:tbls8}-\ref{fig:lrsn8} we show only the results for 8 bytes sized requests which are representative of the behavior  we observed for larger memory requests.
}

In Figures \ref{fig:tbls}-\ref{fig:tbfs} we report data for a comparison among all the back-end allocators implemented in user-space. Here, we configured all the tested allocators to manage chunks of minimal size set to 8 bytes, and maximal size set to 16KB. Also, we used different allocation/deallocation sizes ranging from 8 bytes to 1024 bytes. For the Constant Occupancy test these sizes indicate the minimum ones among those managed, while the maximum ones are set to be 16 times larger.

The data show a clear gain of the non-blocking approach when compared to all the other spin-lock-based solutions especially when increasing the thread count. Also, both the 1-level and 4-level organizations of our non-blocking buddy system have good performance, with the 4-level organization providing a few benefits especially in scenarios where a batch of allocations is interleaved with a batch of releases along the execution of a thread, as in some settings of the Thread Test. Rather, in scenarios where a single allocation is followed by a single release, the lower level of fragmentation of the buddy system allows for less conflicting operations when working with RMW instructions at individual levels, as in the 1-level organization. In any case, when running with 32 threads, the non-blocking buddy system provides performance gain ranging from 9\% to 95\% across all the configurations.

\begin{figure}
\centering
\includegraphics[width=0.8\linewidth]{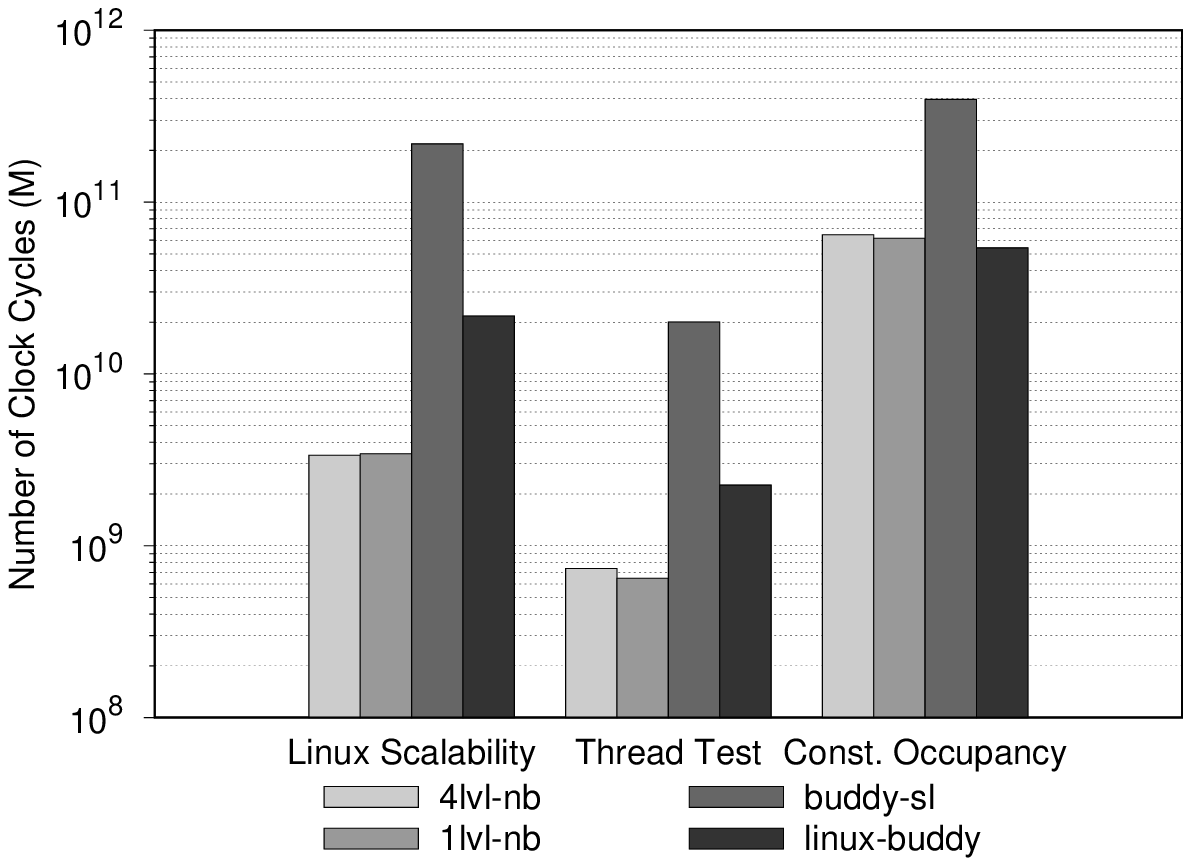}
\vspace{-0.3cm}
\caption{Comparison with the Linux buddy system.}
\label{kernel}
\end{figure}

 In a second set of experiments, whose results are reported in Figure \ref{kernel}, we compare the performance of the different allocators this time including the buddy system from version 3.2 of the Linux kernel---to the best of our knowledge, the buddy system is essentially the same in later versions of the Linux kernel.
 To test with such a kernel level allocator, we developed Linux external modules implementing the logic for the tests related to Linux Scalability, Thread Test and Constant Occupancy.
 This has been done by relying on kernel level threads which interact with the Linux buddy system via {\tt \_\_get\_free\_pages} and {\tt free\_pages} kernel services.
 Given that our target machine has 8 NUMA nodes, and hence the kernel handles 8 instances of a buddy allocator in parallel, to test the access performance to the same allocator instance we set the memory-policy of the threads we activated within the Linux module so as to bind the allocations towards the same buddy-system instance (namely, instance 0).
 We have configured the user-space allocators to manage the same amount of memory as the kernel level buddy allocator, and with the same granularity of minimum and maximum chunk size.
 The data in Figure \ref{kernel} refer to the tests executed when targeting allocations/releases of 128KB chunks (in Constant Occupancy this time the value corresponds to the maximum-size allocatable chunk) with 32 threads.
 Here, we see that our non-blocking version has performance which is comparable with the Linux buddy system in one case---the Constant Occupancy test---and is definitely better with the other two test settings.
 In particular, the non-blocking allocator we propose has performance gain that ranges from 71\% (as in the 1-level case) to  67\% (as in the 4-level case) for Thread Test, and is of the order of 84\% for Linux Scalability.
 

\section{Conclusions}

We have presented a non-blocking approach for back-end memory allocators based on the well-known buddy-system specification.
To date, this is the first practical implementation of a buddy system that jointly supports allocation, deallocation, and coalescing operations, all implemented in non-blocking fashion.
Moreover, we have presented an optimization aimed at reducing\ the number of executed atomic instructions.
The experimental assessment has shown the effectiveness of our solution vs both user-space and kernel-space already-optimized allocators.
As future work, we plan to embed our solution in front-end allocators allowing them to interact more frequently the back-end allocator, thanks to its increased scalability, and to reduce the memory consumption peak.

\newpage



\newpage


\appendix

\section{Progress Guarantees}
In this appendix, we provide a sketch proof that our buddy system achieves lock freedom~\cite{Herlihy2011}, thus guaranteeing that \emph{some} allocation/deallocation invocation eventually completes.
Informally, it is like saying that if a thread does not make progress this is \emph{due to} the advancement (so the progress) in the computation of other threads.
Our proof proceeds in two steps. First, we give a definition of progress in our algorithm, and then we prove that threads are blocked in a retry loop because of conflicts with other threads which anyhow are making progress.

As explained in Section \ref{buddy}, both {\sc NBAlloc} and {\sc NBFree} show (more than) one phase where they try to mark in some way nodes starting from the freed/allocated one to a predefined ancestor (typically the level associated to $max\_level$). We call such phase a $climb$ from a $source$ node to a $destination$ one.

\begin{definition}
A thread performs a step forward (makes progress) in a climb phase if the number of not-yet visited nodes is decreased by one.
\end{definition}

In our code we can identify one loop in the {\sc NBAlloc} procedure and two nested loops in {\sc TryAlloc}, {\sc FreeNode} and {\sc Unmark}.

\begin{lemma}
\label{le:nballociftryalloc}
{\sc NBAlloc} is lock-free iff \, {\sc  TryAlloc} is lock-free.
\begin{proof}
The unique loop in {\sc NBAlloc} is a finite loop where at each iteration local values are computed and two functions are invoked: $is\_free$ which contains no loops and {\sc TryAlloc}.
It follows that {\sc NBAlloc} completes if and only if {\sc TryAlloc} completes.
\end{proof}
\end{lemma}

\begin{lemma}
\label{le:nbfreeiffreenode}
{\sc NBFree} is lock-free iff \, {\sc FreeNode} is lock-free.
\begin{proof}
{\sc NBFree} consists only in invoking the {\sc FreeNode} method.
\end{proof}
\end{lemma}

Analyzing the code of {\sc TryAlloc}, {\sc FreeNode} and {\sc Unmark}, we can see that they contain one occurrence of two nested loops.
These loops have a well-defined structure.
The outer loop performs a climb from a source to a destination, thus it is finite, while the inner loop is a retry cycle for updating node metadata with a CAS.
We refer to such update as a \emph{climb step}.
Consequently, threads can be stuck only in inner loops.

\begin{lemma}
\label{le:progress}
If a thread fails to update node metadata in an inner loop, another thread has made progress.
\begin{proof}
From the CAS semantics, we know that for a thread $A$ failing a CAS on metadata of a node $n$ at level $N$, there is a thread $B$ that has successfully updated such metadata.
Such update can be performed by three operations: (i) T\ref{alg:tryalloc:occupynode}, (ii) F\ref{alg:free:releasenode} or (iii) a climb step.

On one hand, if $B$ executes lines T\ref{alg:tryalloc:occupynode} and F\ref{alg:free:releasenode}, it is clearly making progress, since it will start a new climb phase from the $(N-1)$-th level.
On the other hand, if $B$ completes its inner loop, it makes a step forward along its climb to the destination by moving from the $N$-th level to the ($N-1$)-th level.
\end{proof}
\end{lemma}

\begin{lemma}
\label{le:allocunmarkfreenodelockfree}
{\sc TryAlloc, Unmark, FreeNode} are lock-free functions.
\begin{proof}
Since climbs performed by a function are finite, they are not nested (no threads start a new climb while executing one), and each climb has a finite number of steps, we know from the last lemma that for a thread stuck at level $i$ there is a thread making progress---it succeeds climbing to level $i-1$.
This reasoning can be iterated until there is a thread $A$ stuck while trying to update the destination node of another thread $B$,
 which succeeds and completes its climb.
Since threads performs sequential climbs,
 eventually a thread 
will completes its last climb and, thus, the function execution.
\end{proof}
\end{lemma}

\begin{theorem}
{\sc NBAlloc} and {\sc NBFree} are lock-free functions.
\begin{proof}
The claim follows from Lemma \ref{le:nballociftryalloc}, \ref{le:nbfreeiffreenode} and \ref{le:allocunmarkfreenodelockfree}
\end{proof}
\end{theorem}

\section{Safety Properties}
In this section we define some safety properties intended as ``nothing bad happens'' that we believe an allocator should guarantee and consequently we give an informal proof that our non-blocking buddy system satisfies such properties, 
 described as follows:
\begin{description}
\item[S1.] \label{def:alloc} \emph{A successful allocation returns a non-allocated set of memory addresses coherent respect to the requested size.}
\item[S2.] \label{def:free}  \emph{A correct invocation of a free releases exactly the memory targeted by the request.}
\end{description}

To prove that our non-blocking system satisfies the aforementioned safety conditions, we rely on the following axioms derived by the design of our algorithms:

\begin{description}
\item[AX1:] an {\sc NBAlloc} returns contiguous memory;
\item[AX2:] {\sc NBAlloc} of level $H$ returns a base address $B=k2^H$;
\item[AX3:] {\sc NBAlloc} of level $H$ returns set of addresses $S$ such that $|S|=2^H$;
\item[AX4:] each method execution, that has reached a node $n$ during a climb, has updated the state of all traversed nodes.
\end{description}

\begin{lemma}
\label{le:allocoksize}
A successful allocation returns a memory space coherent respect the size requested.

\begin{proof}
WLOG we consider allocations of size $2^H$. A {\sc NBAlloc} request of size $s$ computes the target level $L$ as $\log_2(s)$ (line A5). Moreover, by AX3 we know that this level contains memory chunks of size $2^L$,
 that is exactly $s$.

In order to successfully complete, an allocation has to find a free node $n$ at level $L$, set its status to fully occupied (line T2), and climb the tree until {\tt max\_level} traversing only free or partially occupied nodes (line T11).
Thus, a successful allocation has set all nodes between the one at level $L-1$ and the one at {\tt max\_level} (from AX4) as partially occupied.

\end{proof}
\end{lemma}

\begin{lemma}
Two different allocated sets of addresses of the same size cannot overlap each other.
\begin{proof}
Let $A=[r_A,r_A+2^H)$ and $B=[r_B,r_B+2^H)$ be two sets of addresses of level $H$ returned by two successful allocations.\\
Supposing that the statement is not true, it follows that
\[r_A < r_B<r_A+2^H\] from AX1 and AX3.\\
We know from AX2 that $r_A=k_A2^H$ and $r_B=k_B2^H$.\\
Since $r_A\not=r_B$ by hypothesis, it follows that $k_A\not=k_B$.\\
WLOG $k_A+D=k_B$, where $k_A,k_B,D\in \mathbb{N}$ and $D>0$.\\
It follows that $A=[2^H k_A,2^H k_A+2^H)$\\ and $B=[2^H (k_A+D),2^H (k_A+D)+2^H)$.\\
It follows that
\begin{gather*}
2^H k_A < 2^H (k_A+D) < 2^H k_A +2^H\\
k_A <k_A+D < k_A +1\\
0 < D < 1
\end{gather*}
which is a contradiction.

\end{proof}
\end{lemma}

\begin{lemma}
Two different node sets are overlapping iff one node is an ancestor of the other.
\begin{proof}
$\Rightarrow$.
Suppose that the statement is not true, namely one node $A$ is not the ancestor of the other $B$ and $A\cap B \neq \emptyset$.

In any case, they have at least common ancestor (e.g. the root). Let $P$ be the lowest level common ancestor at level $l_P$ such that $A \subseteq P$ and $B \subseteq P$.
It follows that $A$ is either a descendant of $P$ or is exactly one child of $P$ and $B$ is either a descendant of $P$ or is the other child of $P$.
As we know, $P$ has two child $L_P$ and $R_P$ such that $L_P \cup R_P = P \wedge L_P \cap R_P = \emptyset$ by construction.
WLOG, we have that $A\subseteq L_P$ and $B\subseteq R_P$.
Consequently,
\begin{gather*}
(A \cap B) \subseteq (L_P \cap R_P)\\
(A \cap B) \subseteq  \emptyset\\
(A \cap B) =  \emptyset
\end{gather*}
which is a contradiction.

$\Leftarrow$.
Let $A$ be a node and let $A_i$ the set of $i$-th descendants of $A$.
It follows,
\[
\bigcup_{a \in A_i} a = A
\]
by construction.
\end{proof}
\end{lemma}

\begin{lemma}
\label{le:allocok}
An alloc never returns an already allocated, but not freed, set of addresses.
\begin{proof}
If the statement is not true, there are at least two successful allocations $A$ and $B$ that have allocated two sets, $S_A$ and $S_B$, of overlapping addresses, namely $S_A\cap S_B \not= \emptyset$.

From Lemmas 5 and 6, we know that either a) $S_A = S_B$ or b) $S_A \neq S_B$ and one node is the ancestor of the other.

Case a)
Both $A$ and $B$ allocations have succeeded to flip the occupancy bit on the same node.
This is impossible unless the node has been freed, which contradicts the hypothesis.

Case b)
WLOG, we assume that $A$ is an ancestor of $B$.
Since both $A$ and $B$ succeeded, they have successfully marked as partially occupied all nodes along their climb to the destination, which is the same for both $A$ and $B$ by construction (line T\ref{alg:tryallocupperbound}).
It follows that $B$ has successfully set one of the partial occupancy bits ({\tt OCC\_LEFT} or {\tt OCC\_RIGHT}) of the node $n_A$ occupied by $A$.
Moreover, $A$ has successfully set the occupancy bit of $n_A$.
The set of the occupancy bit is performed with a CAS using 0 as old value, thus $A$ makes the occupancy bit flip from 0 to 1 ($00000 \rightarrow 10011$).
On the other hand, partially occupancy bits are only set ($0\_\_00 \rightarrow 0\_\_11$) if and only if the occupancy bit is not set (line T\ref{alg:tryalloc:ifoccreturn}).
Since both CASes are atomic, it means that bits have assumed the following values in time:
\[
00000 \underbrace{\rightarrow 10011}_{\text{applied by $A$}} \xrightarrow[]{*} 0\_\_00  \underbrace{\rightarrow 00011}_{\text{applied by $B$}}.
\]
It follows that a transition $1 \rightarrow 0$ of the occupancy bit has taken place.
This transition is applied only at line F\ref{alg:free:releasenode} of {\sc FreeNode}, while releasing a particular node passed has parameter.
Since a successful allocation does not ever call a {\sc FreeNode} during its last iteration, it cannot be applied by $A$ or $B$.
It follows that a third operation $C$ has reset the occupancy bit of $n_A$.
In particular, $C$ is a {\sc FreeNode} invocation on $n_A$ that contradicts the hypothesis that both sets allocated by $A$ and $B$ are not freed.
\end{proof}
\end{lemma}

\begin{theorem}
A successful {\sc NBAlloc} satisfies S1.

\begin{proof}
It follows from Lemmas \ref{le:allocoksize} and \ref{le:allocok}.
\end{proof}
\end{theorem}

\begin{definition}
A free \emph{f} is invoked \emph{correctly} iff the allocation of the memory to be released has returned before the invocation of \emph{f}.
\end{definition}

\begin{lemma}
\label{le:freeoksize}
An invocation of a free releases the memory  targeted by the free request.

\begin{proof}
A free operation translates the memory address $a$, passed as argument to {\sc NBFree}, to the index $n$ of the relative node. Since this index is computed by applying a transformation that is the inverse respect the one applied by the {\sc NBAlloc} to compute the address to be returned, we are sure that the node to be free is the same allocated by the method that has returned the memory address $a$. Since a {\sc NBFree} procedure completes only when it finds a node partially occupied by other allocations (line U8), we are sure that all the traversed nodes have been set to 0 from AX4.
\end{proof}
\end{lemma}

\begin{lemma}
\label{le:freeok}
A correct invocation of a {\sc NBFree} does not release a memory space different by the one targeted by the request.

\begin{proof}
Considering, by absurd, that the former statement is not true, we have that a thread A performing a {\sc NBFree} $F_A$ has successfully updated via a CAS (line U12) a node $n$, notifying that the relative subtree (the one where $F_A$ operation is started) is totally released when a sub-portion of this tree is still allocated --- if this happens, the state of the node $n$ could transit in a released state (all status bits set to 0) and then it can serve another allocation. By construction (lines U8-U10), if A has reached the node $n$ during an unmarking climb, it means that it has seen the relative son $s$ of this node with a state equal to 0. Then, a portion of memory in the subtree could be allocated only if there is a concurrent thread B executing a {\sc NBAlloc} that has updated the node $s$ after A.

 At this point there are two possibilities: B updates the node $n$ after A; B updates the node $n$ before A.
In the first case, since the thread B will update in turn the state of the node, the absurd is not in place (the free appears to precede the allocation).
In the second case, the thread A has reset the relative coalescing bit of $n$ (lines T15-T17), but this is absurd since A can succeed updating $n$ only if the relative coalescing bit is set to 1.
\end{proof}
\end{lemma}

\begin{theorem}
A correct invocation of \, {\sc NBFree} satisfies S2.
\begin{proof}
It follows from Lemma \ref{le:freeoksize} and \ref{le:freeok}.
\end{proof}
\end{theorem}

\end{document}